\documentclass[prb,nobibnotes,floatfix,superscriptaddress,twocolumn,showpacs]{revtex4-1}
\usepackage{amsfonts,amsmath,graphicx,epsfig,latexsym}
\usepackage{subfigure}
\usepackage{graphicx}
\usepackage{color}
\usepackage{natbib}
\usepackage{multirow}
\usepackage{latexsym,amssymb}
\usepackage{amsmath}
\usepackage{soul}

\begin{document}
\title{ Valence and spin fluctuations in the Mn-doped ferroelectric BaTiO$_3$}
\author{Subhasish Mandal}
\affiliation
	{Department of Applied Physics, Yale University, New Haven, Connecticut  06511, USA}
\affiliation
{Extreme Materials Initiative, Geophysical Laboratory, Carnegie Institution for Science, Washington D.C. 20015, USA}

\author{R. E. Cohen}
\affiliation
{Extreme Materials Initiative, Geophysical Laboratory, Carnegie Institution for Science, Washington D.C. 20015, USA}
\affiliation
{Department of Earth and Environmental Sciences, LMU Munich,  80333 Munich, Germany}
\author{K. Haule}
\affiliation  
{Department of Physics, Rutgers University, Piscataway, New Jersey 08854, USA}

\begin{abstract}
{\footnotesize
We study Mn substitution for Ti in BaTiO$_3$ with and without compensating oxygen vacancies using density functional theory (DFT) in combination with dynamical mean field theory (DMFT). We find strong charge and spin fluctuations.  Without compensating oxygen vacancies,  the ground state is found to be a quantum superposition of two distinct atomic valences, 3{\it d}$^4$  and 3{\it d}$^5$.  Introducing a compensating oxygen vacancy at a neighboring site reduces both charge and spin fluctuations due to the reduction of electron hopping from Mn to its ligands.  As a consequence, valence fluctuations are reduced, and the valence is closely fixed to the high spin 3{\it d}$^5$ state.  Here we show that inclusion of charge and spin fluctuations is necessary to obtain an accurate ground state of transition metal doped ferroelectrics. 
}
\end{abstract}


\maketitle
\newpage
\section{Introduction}   
Multiferroics respond to both electric and magnetic fields, and their coupling is an exciting field of research for both understanding their fundamental physics and for potential device applications. One promising route to magnetoelectric materials is to dope ferroelectrics with magnetic ions\cite{nmat1,nmat2,Norton01022003,PhysRevB.87.214110}. Such materials are of interest for electronics that can integrate data processing and memory operation in a single solid state device\cite{nmat1,nmat2,JAP1,JAP2}. Many commercial transducer materials are doped with transition metal impurities to improve piezoelectric properties, mechanical quality factor, and coercive field, and to decrease electrical conductivity \cite{aip1,0957-4484-17-5-020,0957-4484-17-5-020,0953-8984-20-50-505209,0953-8984-20-8-085206,doi:742066,Dang,doi:1499359,doi:1511757,doi:1812576,nmat1,nmat2,JAP1,JAP2,Norton01022003}, but the exact role of the doped impurities is unclear. Defect dipoles formed by transition metal dopants with oxygen vacancy neighbors can greatly enhance electromechanical coupling \cite{Ren2004,Zhang2005,Nossa,Chapman2017,Liu2017}. Thus, the electronic structure of transition metal dopants is of great interest in general, and in particular in multiferroics and dilute magnetic semiconductors \cite{RN14480,RN14521}.

When a transition metal ion is doped into a classic ferroelectric material like BaTiO$_3$, whether the impurity would be an acceptor or a donor of electrons depends on the number of valence electrons and  3{\it d} occupation, i.e oxidation state.   Valence, charge, or oxidation state are concepts commonly used in chemistry. Oxidation state is an ill-defined quantity in quantum mechanics, although it has proven extremely useful in chemical intuition  \cite{pauling,PhysRevLett.109.216401,PhysRevLett.108.166403,car1,pk2,PhysRevB.86.134503,PhysRevB.96.165135}. Oxidation state can be a point of confusion as very often the charge or oxidation state of a cation differs significantly from the Born transverse effective charges, or the static charges computed  by projection onto local or Wannier orbitals or from orbital occupations \cite{PhysRevLett.109.216401,PhysRevLett.108.166403,car1,pk2,PhysRevB.86.134503}. The {\it d}-occupation remains invariant in charge order driven metal-insulator transitions\cite{PhysRevLett.109.216401}, but does depend on the choice of orbitals. The {\it d} occupation is also related to the ion magnetic moment.  

Mn commonly has three different oxidation states  (Mn$^{4+}$,  Mn$^{3+}$,  Mn$^{2+}$) in perovskites. Mn$^{4+}$ and  Mn$^{2+}$ have an electronic configuration of {\it d}$^3$ and {\it d}$^5$ respectively, or half-filled t$_{2g}$ or t$_{2g}$+e$_g$ manifolds in octahedral symmetry. Which states of valence the paramagnetic ions are incorporated into the BaTiO$_3$ and other perovskites is an open and long standing problem \cite{aip1,0957-4484-17-5-020,Dang,0953-8984-20-50-505209,0953-8984-20-8-085206,doi:742066,Dang,doi:1499359,doi:1511757,doi:1812576}. The magnetic moment of Mn substituting for Ti on the B-site in Mn doped BTO without any compensating oxygen vacancy is  3 $\mu_B$ (Mn$^{4+}$)\cite{lda1,Nossa} in conventional DFT and DFT+U.  Electron paramagnetic resonance (EPR)  and X-ray photoelectron spectroscopic measurements on Mn doped BTO show that Mn can exist in various charge states in BTO; some EPR measurements performed on Mn doped BTO nanoparticle show a high spin state of Mn with the moment of 5  $\mu_B$ (Mn$^{2+}$)\cite{JAP2}, whereas Mn$^{4+}$ with 3$\mu_B$  is also found\cite{aip1}. 
The oxidation state of the Mn depends on the oxygen fugacity during growth or annealing, and depends on the concentration of compensating oxygen vacancies or other impurities and defects. Using density functional calculations, Nossa {\it et al.} found that depending on the oxygen vacancy, 
Mn ions in BaTiO$_3$ can exist either on high spin state ( Mn$^{2+}$ ) or low spin state (Mn$^{4+}$)\cite{Nossa}.  

  In this paper, we carefully investigate electronic structure of Mn doped BTO with and without compensating O-vacancies in the paramagnetic phase of Mn. Using a {\it state-of-the-art} DFT+DMFT method, we focus on understanding the  charge and spin states Mn exhibits in these prototypical systems in order to combine both $d^0$ states and partially occupied  {\it d} -states to unite ferroelectricity and magnetism in one material. We compare our results with conventional DFT and DFT+U in the ferromagnetic phase of Mn. We then describe the effects of oxygen vacancies on the charge and spin fluctuations of Mn and explore the change in local magnetic moment of Mn. 
 
\section{ Methods and Structural details  }

To understand better the electronic structure of transition metal dopants in dielectrics, in general, and Mn$_{Ti} \pm $V$_O$ in BaTiO$_3$ in particular, we use dynamical mean-field theory (DMFT) \cite{RevModPhys.78.865}, a sophisticated method, which includes quantum dynamical effects, and takes into account both valence and spin fluctuations. In contrast, DFT includes only average interactions, and DFT+U includes only a single configuration, ignoring multiplet effects.
 DFT+DMFT has been very successful in describing strongly correlated materials like high temperature superconductors,  Mott insulators and several transition metal bearing compounds \cite{Shim,peng,haule3,haule_spin,Wang:2013gz,Schafgans,PhysRevB.89.220502,PhysRevB.90.060501,PhysRevLett.119.067004,nmat_kunes}.  In DFT+DMFT, the self-energy that samples all local skeleton Feynman diagrams is added to the DFT Kohn-Sham Hamiltonian \cite{RevModPhys.78.865,nominal1}.   This implementation is fully self-consistent \cite{nominal1,haule3}. 
The iterations stop after full convergence of the charge density, the impurity level, the chemical  potential, the self-energy, and the lattice and impurity Green's functions. The lattice is represented using the full potential linear augmented plane wave (LAPW) method, implemented in the Wien2k \cite{wien2k} package in its generalized gradient approximation (Wu-Cohen-GGA) \cite{PhysRevB.73.235116}. The continuous time quantum Monte Carlo method is used to solve the quantum impurity problem and to obtain the local self-energy due to the correlated Mn 3{\it d} orbitals. The self-energy is analytically continued from the imaginary to real axis using an auxiliary Green's function to obtain the partial density of states. 
 A fine k-point mesh of  at least 4$\times$ 4 $\times$ 4 k-points in Monkhorst-Pack k-point grid and a total 40 million Monte Carlo steps for each iteration are used for the paramagnetic phase of the Mn doped BTO at T=300K. 
The Coulomb interaction $U$ and Hund's coupling $J_{H}$  are fixed at 6.0 eV and 0.8 eV, respectively \cite{Kutepov:2010bu}, and we have tested varying these parameters. We use the fully localized limit (FLL) double counting \cite{FLL}, as well as ``exact" double counting\cite{exact}. For DFT and DFT+U, we use the all electron LAPW method as implemented in {\sc Wien2k}\cite{wien2k}.  The same $U$ and $J_{H}$  of 6.0 eV and 0.8 eV are used for DFT+U computations respectively. 

 We study a paramagnetic Mn dopant in a supercell with  (Mn$_{Ti}$V$_O$) and without (Mn$_{Ti}$) a neighboring compensating oxygen vacancy using DFT-DMFT \cite{url1} computation at room temperature. 
Two  2$\times$2$\times$2  supercell structures are considered here; one structure with one Mn-replacing Ti atom, and the other structure is with Mn-replacing Ti with an Oxygen vacancy next to the Mn atom along c-axis. These structures were optimized using DFT+U as implemented in ABINIT \cite{Gonze20092582,gonze2005brief} and also used in Ref.\cite{Nossa} to understand the role of Mn doping in BTO. For DFT computations, we consider ferromagnetic order with a single Mn-atom in the supercell.

\section{Results and Discussions}

\begin{figure}
\includegraphics[width=210pt, angle=0]{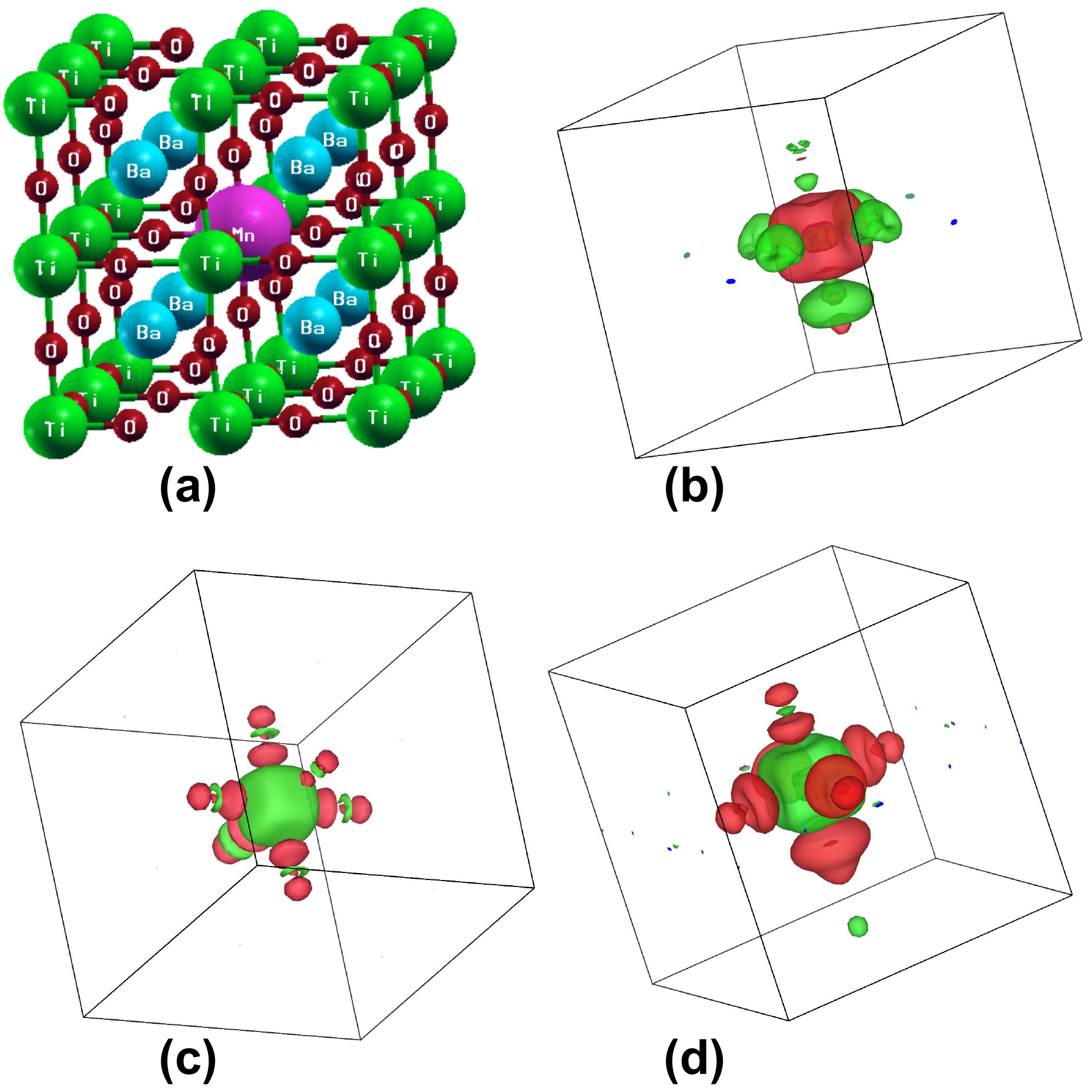}
\caption{ (Color online)
(a) Schematic representation of the structure for  Mn-doped (at Ti-site) BaTiO$_3$  without any compensating oxygen vacancy.  Isosurface plot of electron density difference between (b) DFT and DFT+ U (c) DFT and DFT+DMFT and (d) DFT+U and DFT+DMFT methods. The green or red  means increase or decrease of 0.85 $\times$ $10^{-3}$ e / $\AA^3$  upon the DMFT calculation. 
}
\end{figure}

\begin{table*}
\begin{tabular}{lcccccr} 
\hline
 Method & System &   Magnetic Moment ($\mu_B$) &  Occupation & Band Gap (eV)\\    \hline
NM-DFT  &Mn$_{Ti}$  &  -- & 4.65  &0.00\\  
SP-DFT  & Mn$_{Ti}$  &  3.00 & 3.16(up),0.87(dn)  &1.46\\ 
DFT+U   & Mn$_{Ti}$  &  3.00 & 3.42(up), 1.07 (dn) &1.70\\ 
DFT+DMFT  & Mn$_{Ti}$  &  2.90 & 4.44 &1.14\\ 
NM-DFT & Mn$_{Ti}V_{O}$ & --& 4.79 & 0.00 \\  
SP-DFT   & Mn$_{Ti}V_{O}$ &  4.47 & 3.78(up), 0.51 (dn) &0.00\\  
DFT+U   & Mn$_{Ti}V_{O}$ &  5.00 & 3.34(up), 0.34 (dn) &1.15\\ 
DFT+DMFT  & Mn$_{Ti}V_{O}$  &  4.05 & 4.80 &1.80\\  
\end{tabular}
\caption{Computed magnetic moment  (in $\mu_B$), occupation of Mn {\it d}-orbital, and band gap(eV) for  Mn in BaTiO$_3$  with and without O-vacancy obtained within non-magnetic (NM) DFT, spin-polarized(SP) DFT with ferromagnetic order, DFT+U and DFT+DMFT methods. Here DFT+DMFT is performed in paramagnetic phase of the materials and the magnetic moments in DFT+DMFT represent the average fluctuating local moment; the occupations in DFT and DFT+U are obtained by integrating the projected DOS to the Fermi energy.}
\end{table*}

\begin{center}
\begin{figure*}
\includegraphics[width=480pt, angle=0]{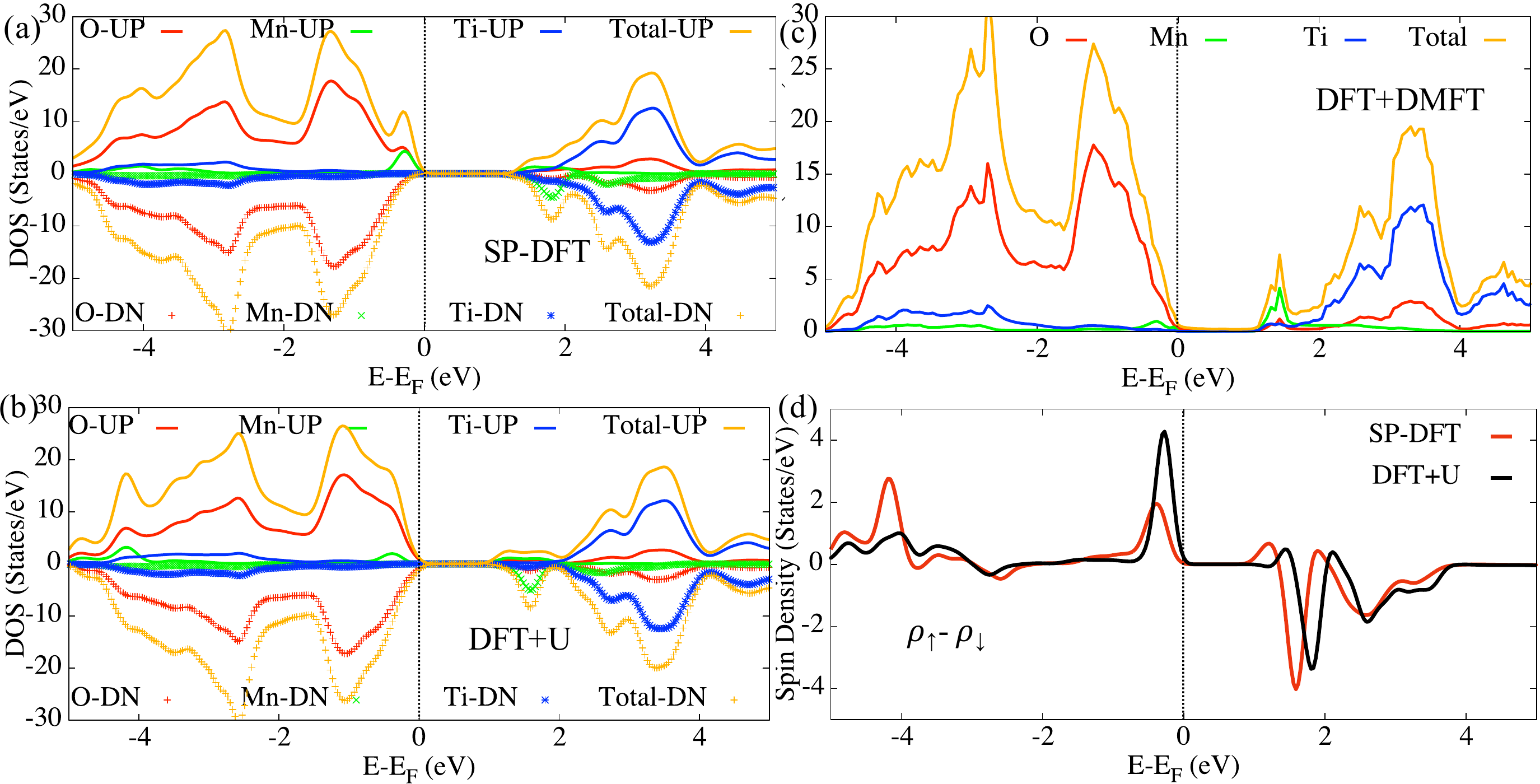}
\caption{(Color online).
Spin decomposed total and projected densities of states (DOS) computed in (a) DFT, (b) DFT+U, and (c) DFT+DMFT methods for Mn substitution of Ti in BaTiO$_3$ without any compensating oxygen vacancy ( Mn$_{Ti}$);  (d) Computed spin density ($\rho_\uparrow - \rho_\downarrow$) with DFT and DFT+U (d) for the same system. 
} 
\end{figure*}
\end{center}

\begin{figure*}
\includegraphics[width=480pt, angle=0]{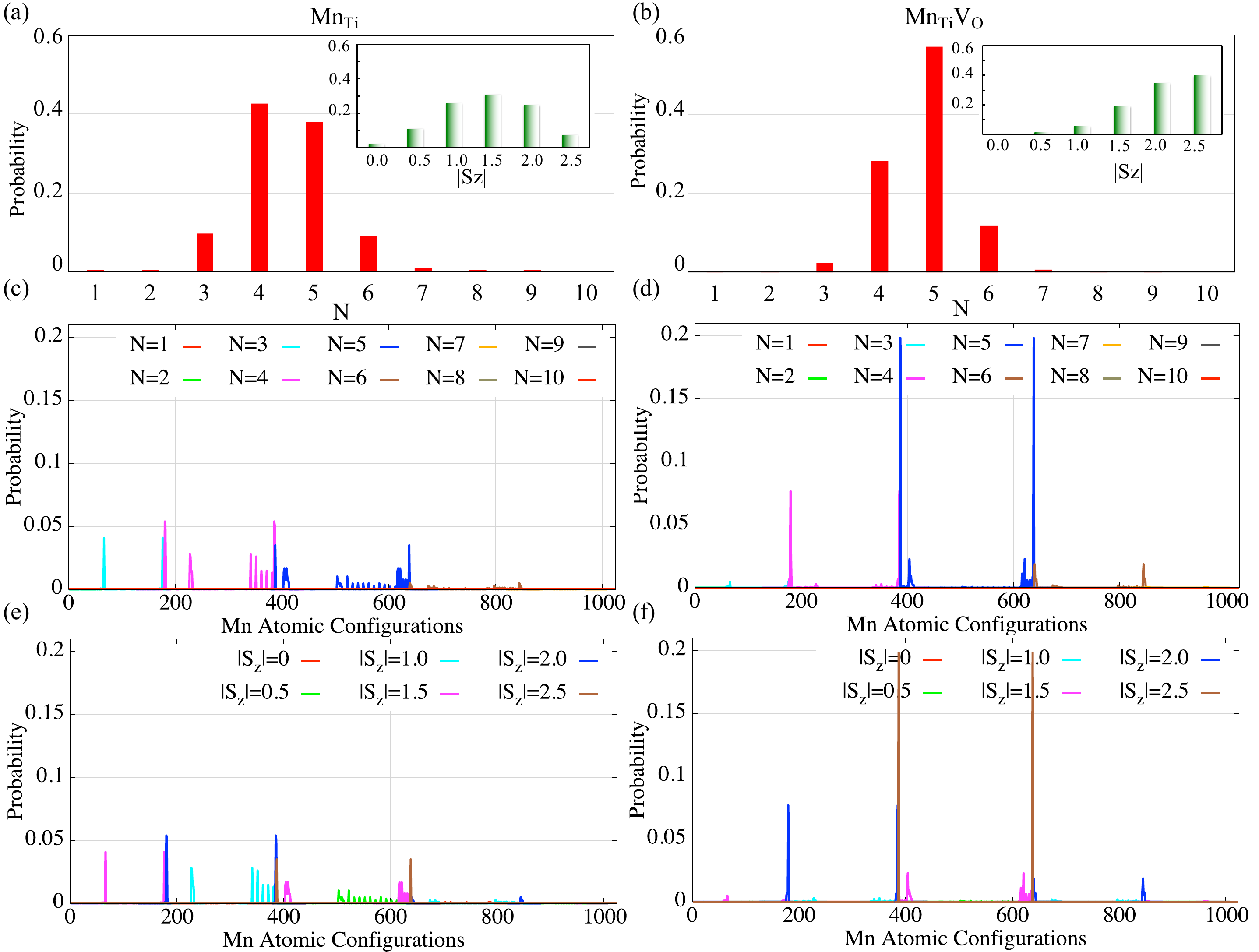}
\caption{(Color online).
DFT+DMFT computed atomic histogram of the Mn 3{\it d} shell for Mn substituted BaTiO$_3$ (left) without (Mn$_{Ti}$) and (right) with (Mn$_{Ti}$V$_O$) compensating oxygen vacancy:  (a -b)decomposed in number of particles N and (inset) spin-state (S$_Z$);  probability distribution for all 2$^{10}$(=1024) atomic configurations sorted for each N (c-d) and for each spin-state (e-f);  N and S$_Z$ values are denoted with various colors.   }
\end{figure*}

Firstly, we describe the effects of a Mn impurity without any compensating oxygen vacancy (Mn$_{Ti}$). The charge density $\rho(r)$ is computed using spin-polarized DFT, DFT+U, and  DFT+DMFT (Fig. 1). The difference in $\rho(r)$ in DFT and DFT+U shows that DFT places more charge on the Mn atoms compared to DFT+U (Fig1b).  DFT+DMFT always places more charge on the Mn atom than other two methods. 
 
To better understand these differences, we compare the total and the partial densities of states (DOS) for spin-polarized DFT, DFT+U, and DFT+DMFT (Fig. 2). The partial DOS shows that most of the contribution around E$_F$ is from O-atoms  (Fig. 2a-c).  
 The 3{\it d} -occupation is consistent with Mn$^{4+}$ in both DFT and DFT+U, as found in previous DFT-based studies \cite{lda1,Nossa} (Table I).  Although the non-magnetic DFT predicts the occupation to be 4.65, it puts the Mn {\it d} orbital at E$_F$, and results a metallic solution \cite{lda1}. 
 The magnetic moment in DFT+U is 2.67 $\mu_B$  inside the Mn muffin-tin sphere,  0.24 $\mu_B$ for the interstitial, and  0.09 $\mu_B$ inside the other spheres, giving a
total moment of 3.00 $\mu_B$ (Table I). The equal magnetic moment in DFT and DFT+U can also be noticed from from the integration of the spin-density (Fig. 2d) of the {\it d} orbital until E$_F$.
 
The computed Mn 3{\it d} occupation in DFT+DMFT is 4.44, and the average fluctuating local moment is 2.90$\mu_B$. We compare our results with different double counting (DC) schemes; a fully localized limit form of double counting or FLL as introduced in Ref. \cite{FLL}, ``nominal" DC as introduced in \cite{nominal1,nominal2}  and ``exact" double counting \cite{exact}. We use both density-density form of the Coulomb repulsion (Ising) and the full Coulomb repulsion (Full) in rotationally invariant Slater form (Table II) \cite{coulomb2}.   
Changing U does not affect the {\it d}-occupation significantly.  Keeping J$_{H}$=0.8 eV and varying U from 2 eV to 6 eV, the occupation changes from 4.74 to 4.78. Also keeping U fixed at 6 eV and varying J$_{H}$ from 0 to 1.2 eV, the occupation changes from 4.30 and 4.40 and reaches its maximum value of 4.77 for J$_{H}$=0.8 eV.

We  next discuss the results for Mn doped BTO with a compensating oxygen vacancy along the z-direction (Mn$_{Ti}$V$_O$) \cite{Nossa}.  
Interestingly, the computed occupancy of Mn$_{Ti}$ -V$_O$ is found to be 3{\it d}$^5$ in either DFT, DFT+U, or DFT+DMFT (Table I).  For DFT and DFT+U, we find the moment to be 4.47 and 5.00 $\mu_B$  respectively for the entire cell. 
In DFT+U, the moment in the Mn muffin-tin sphere is 4.28 $\mu_B$, in the interstitial is 0.51 $\mu_B$, and the other spheres have a moment of 0.21 $\mu_B$, giving a total of 5.00 $\mu_B$ for the cell.
 Whereas the magnetic moment computed in DFT and DFT+U are for the ferromagnetic phase, for DFT+DMFT we compute the average fluctuating moment ($<m_z>$) in the paramagnetic phase of Mn using the formula : $<m_z>=2 \times \sum_i P_i \times |S_z|_i$, where $P_i$ is the probability of the $i^{th}$ multiplet in the CTQMC impurity solver and 
$|S_z|_i$ is the absolute value of the corresponding moment. In DFT+DMFT the magnetic moment is mostly concentrated on the magnetic ion; the induced moment on the interstitial charge or other non-magnetic ions in the unit cell is very small. Hence we can compare DFT+DMFT local moment with the total moment of the cell within DFT and DFT+U calculations (table I).

\begin{table}
\begin{tabular}{lclcr} 
\hline
 DC   & Coulomb   & n$_d$   \\    \hline
Exact  &   Ising  &   4.44\\  
 Exact &   Full &  4.39  \\  
 Nominal & Ising & 4.75 \\ 
 FLL & Full &4.76 \\ 
 FLL & Ising &4.77 \\ 
\end{tabular}
\caption{DFT+DMFT computed  occupation of {\it d}-orbital in Mn$_{Ti}$ using different double counting and Coulomb schemes\cite{exact,FLL,nominal1,nominal2,coulomb}.}
\end{table}

We now discuss the valence fluctuations of Mn in DFT+DMFT, where the ground-state wave function is not restricted to being a single multiplet, as in DFT+U. On a single atom, there are 1024 different possible multiplets for {\it d}-electrons, characterized by different valences, orbital occupations, and spins \cite{haule3}. The histograms in Fig. 3 describe the probability of finding a Mn atom in the solid in each multiplates, and show that any method that considers only a single multiplet, such as even single determinant group state Quantum Monte Carlo will only be approximate.
We find many occupied multiplets, such as 3{\it d}$^3$, 3{\it d}$^4$ and  3{\it d}$^5$ with a maximum occupation of 3{\it d}$^4$  for Mn$_{Ti}$ and  3{\it d}$^5$  for Mn$_{Ti}$-V$_O$ (Fig. 3a-b). 
Without compensation, we find only about 5\% weight in the most probable configurations. With a compensating vacancy, the fluctuations are smaller, but still there is only about 20\% weight in the most likely multiplets. In Mn$_{Ti}$ the sum of the probabilities are found to be 0.43, 0.34, 0.09 respectively for N=4, 5, and 6. Thus the system is in a mixed valence state with an average {\it d} occupation of $\sim$ 4.4. For Mn$_{Ti}$-V$_O$, the sum of the probabilities are 0.28, 0.57, 0.11 for N=4, 5, and 6 respectively. 
 The difference in probabilities between N=4 and N=5 reduces with compensating oxygen vacancy in Mn$_{Ti}$V$_O$ (Fig. 3b); the probability for N=5 increases and becomes the most probable state. This leads to an increase in average 3{\it d} occupation from 4.4 to 4.8 with compensating O-vacancy.  

\begin{figure}
\includegraphics[width=240pt, angle=0]{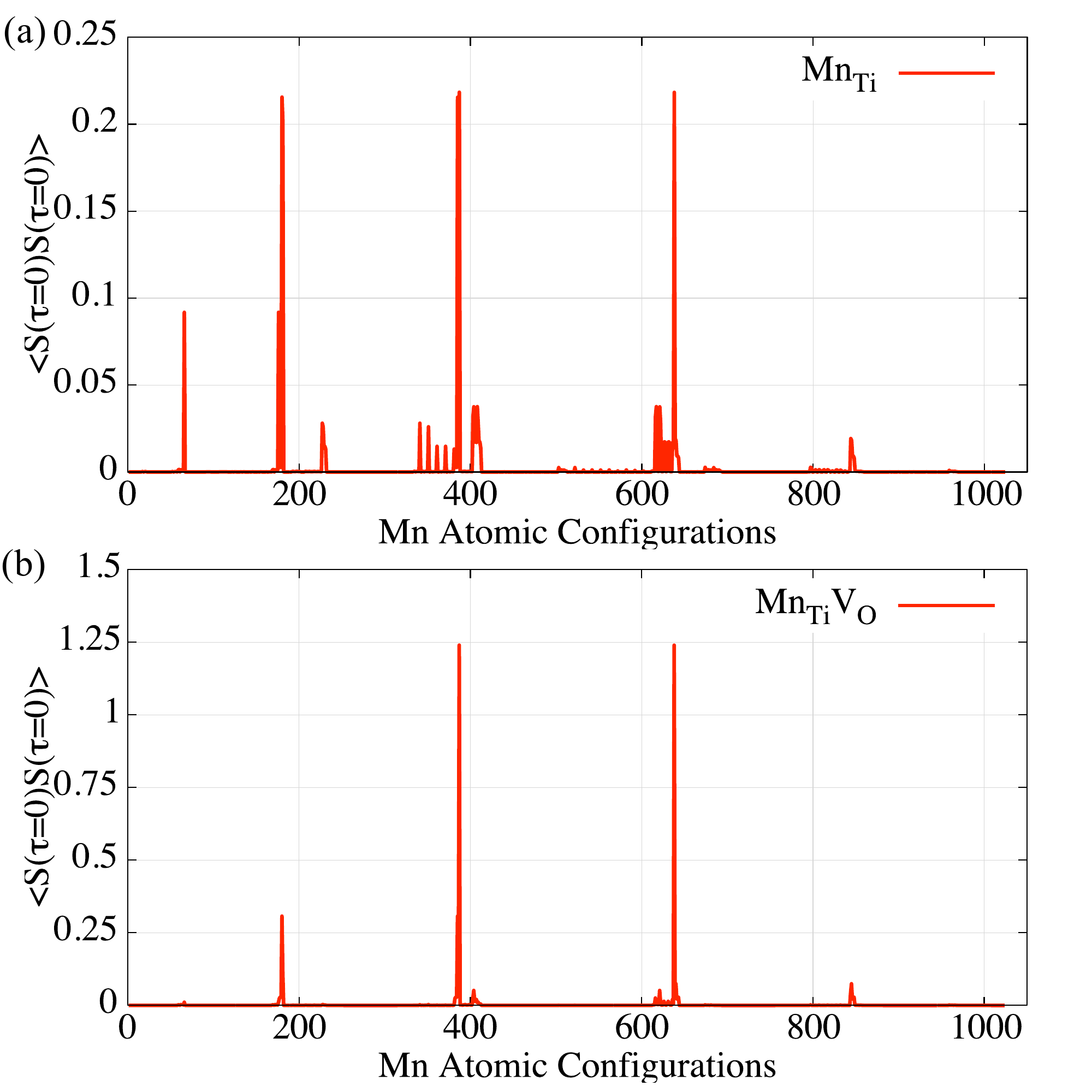}
\caption{ (Color online)
 DFT+DMFT computed squared fluctuating local moment or the averaged spin-spin correlation function (a) for Mn substituted BaTiO$_3$ without (Mn$_{Ti}$) and (b) with compensating oxygen vacancy (Mn$_{Ti}$V$_O$).
}
\end{figure}
The histograms of the CTQMC (Inset of Fig. 3a-b) show the largest probability for the spin state S$_z$=1.5 for Mn$_{Ti}$ and S$_z$=2.5 for Mn$_{Ti}$V$_O$. The histograms also show strong spin fluctuations. 
To identify the associated spin state for each eigenstate, we present them in various colors in Fig. 3(e-f).  Here the first (last) few states with a particular N show the high (low) spin state (Fig. 3c-f). For Mn$_{Ti}$,  we clearly see the spikes in probability for the high spin states (S$_z$=  2, 2.5) at the beginning of the constant N interval as well as for the low spin states (S$_z$= 0, 0.5), at the end of the constant N interval (Fig. 3c and 3e).  For Mn$_{Ti}$V$_O$ we see spikes in probability for high spin states  (S$_z$=2, 2.5) are dominated than that for the low spin states (S$_z$= 0,0.5) (Fig. 3d and 3f) for N=4 and N=5. The overall probability distribution changes significantly with oxygen vacancy and peaks mainly for N=5 and S$_z$= 2.5. Thus both the charge and spin fluctuations are reduced in Mn$_{Ti}$V$_O$ since the hopping of  the electrons from Mn to neighboring oxygen is reduced due to the compensating  vacancy.  
The dominance of the S$_z$ = 2.5  configuration implies that Mn in BTO has a predominantly high spin state in Mn$_{Ti}$V$_O$. The increase in magnetic moment with O-vacancy, as found in Table I, can be explained from the DFT+DMFT computed time averaged spin-spin correlation function ($<$S($\tau$=0)S($\tau$=0)$>$) or the squared fluctuating local moment (Fig. 4), where we find an increase in the local moment in Mn$_{Ti}$V$_O$.  Reduced fluctuation and increased probability for electron to spend more time in the high spin states with O-vacancy gives rise to the increase in  average fluctuating local moment.    

\section{Conclusions} 

Unlike previous density functional based studies yielding either high ( Mn$^{2+}$ ) or low spin ( Mn$^{4+}$ ) state of Mn in BaMn$_{1-x}$Ti$_{x}$O$_{3}$ with and without compensating oxygen vacancy respectively, DFT+DMFT predicts a mixed valence state of Mn in either system. 
Without compensating oxygen vacancies, the ground state in DFT+DMFT is found to be a quantum superposition of two distinct atomic valences, 3{\it d}$^4$  and 3{\it d}$^5$.
Introducing a compensating oxygen vacancy at a neighboring site of Mn reduces hopping of electron from Mn to its ligand. This results in reduction of  both charge and spin fluctuations. We conclude that the charge and valence fluctuations in 3{\it d}-transition metal doped BTO are strong, and are not captured by conventional DFT or DFT+U.  DFT+DMFT predicted average valence of Mn in BTO is 3{\it d}$^{2+}$  and 3{\it d}$^{3.5+}$ with and without compensating oxygen vacancy respectively.
The most important result presented here are the extreme fluctuations predicted, with only a small fraction in any given multiplet, and very different from  DFT+U which assumes a single multiplet to be occupied. Even very accurate methods such as diffusion Monte Carlo generally consider only a single Slater determinant for solids.  DFT+DMFT gives a new insight into the fluctuating states in correlated solids, which is presently challenging to represent by other standard methods.  Our predictions of valence fluctuations in transition metal doped ferroelectrics can be verified, as was done in heavy Fermion system through hard X-ray photoemission study \cite{PhysRevLett.104.247201}, in topological Kondo insulators through muon spin relaxation studies \cite{topo1,topo2} and in perovskite through Mossbauer spectroscopy \cite{doi:10.1063/1.126518} and NMR \cite{PhysRevB.89.174412}.  
\section{Acknowledgement}
 
This research was supported by the office of Naval Research (ONR) grants N00014-12-1-1038 and N00014-14-1-0561 and by the  Carnegie Institution of Washington. S. M acknowledges the support from the NSF SI2-SSI program (Grant No. ACI-1339804).  K. H acknowledges the supports from NSF DMR-1709229. REC is supported by the Carnegie Institution and by the European Research Council advanced grant ToMCaT. Computations were performed at the `Supermike' at the Louisiana State University and NERSC supercomputing facility. We thank Sohrab Ismail-Beigi, Ivan I. Naumov and Javier F. Nossa for helpful discussions to prepare this manuscript. We acknowledge Peng Zhang, Mark Jarrell and Juana Moreno for their important help.

\bibliography{super}

\begin{thebibliography}{58}%
\makeatletter
\providecommand \@ifxundefined [1]{%
 \@ifx{#1\undefined}
}%
\providecommand \@ifnum [1]{%
 \ifnum #1\expandafter \@firstoftwo
 \else \expandafter \@secondoftwo
 \fi
}%
\providecommand \@ifx [1]{%
 \ifx #1\expandafter \@firstoftwo
 \else \expandafter \@secondoftwo
 \fi
}%
\providecommand \natexlab [1]{#1}%
\providecommand \enquote  [1]{``#1''}%
\providecommand \bibnamefont  [1]{#1}%
\providecommand \bibfnamefont [1]{#1}%
\providecommand \citenamefont [1]{#1}%
\providecommand \href@noop [0]{\@secondoftwo}%
\providecommand \href [0]{\begingroup \@sanitize@url \@href}%
\providecommand \@href[1]{\@@startlink{#1}\@@href}%
\providecommand \@@href[1]{\endgroup#1\@@endlink}%
\providecommand \@sanitize@url [0]{\catcode `\\12\catcode `\$12\catcode
  `\&12\catcode `\#12\catcode `\^12\catcode `\_12\catcode `\%12\relax}%
\providecommand \@@startlink[1]{}%
\providecommand \@@endlink[0]{}%
\providecommand \url  [0]{\begingroup\@sanitize@url \@url }%
\providecommand \@url [1]{\endgroup\@href {#1}{\urlprefix }}%
\providecommand \urlprefix  [0]{URL }%
\providecommand \Eprint [0]{\href }%
\providecommand \doibase [0]{http://dx.doi.org/}%
\providecommand \selectlanguage [0]{\@gobble}%
\providecommand \bibinfo  [0]{\@secondoftwo}%
\providecommand \bibfield  [0]{\@secondoftwo}%
\providecommand \translation [1]{[#1]}%
\providecommand \BibitemOpen [0]{}%
\providecommand \bibitemStop [0]{}%
\providecommand \bibitemNoStop [0]{.\EOS\space}%
\providecommand \EOS [0]{\spacefactor3000\relax}%
\providecommand \BibitemShut  [1]{\csname bibitem#1\endcsname}%
\let\auto@bib@innerbib\@empty
\bibitem [{\citenamefont {Efremov}\ \emph {et~al.}(2004)\citenamefont
  {Efremov}, \citenamefont {van~den Brink},\ and\ \citenamefont
  {Khomskii}}]{nmat1}%
  \BibitemOpen
  \bibfield  {author} {\bibinfo {author} {\bibfnamefont {D.~V.}\ \bibnamefont
  {Efremov}}, \bibinfo {author} {\bibfnamefont {J.}~\bibnamefont {van~den
  Brink}}, \ and\ \bibinfo {author} {\bibfnamefont {D.~I.}\ \bibnamefont
  {Khomskii}},\ }\href@noop {} {\bibfield  {journal} {\bibinfo  {journal} {Nat.
  Mater}\ }\textbf {\bibinfo {volume} {3}},\ \bibinfo {pages} {853 } (\bibinfo
  {year} {2004})}\BibitemShut {NoStop}%
\bibitem [{\citenamefont {Ederer}\ and\ \citenamefont {Spaldin}(2004)}]{nmat2}%
  \BibitemOpen
  \bibfield  {author} {\bibinfo {author} {\bibfnamefont {C.}~\bibnamefont
  {Ederer}}\ and\ \bibinfo {author} {\bibfnamefont {N.~A.}\ \bibnamefont
  {Spaldin}},\ }\href@noop {} {\bibfield  {journal} {\bibinfo  {journal} {Nat.
  Mater}\ }\textbf {\bibinfo {volume} {3}},\ \bibinfo {pages} {849 } (\bibinfo
  {year} {2004})}\BibitemShut {NoStop}%
\bibitem [{\citenamefont {Norton}\ \emph {et~al.}(2003)\citenamefont {Norton},
  \citenamefont {Theodoropoulou}, \citenamefont {Hebard}, \citenamefont
  {Budai}, \citenamefont {Boatner}, \citenamefont {Pearton},\ and\
  \citenamefont {Wilson}}]{Norton01022003}%
  \BibitemOpen
  \bibfield  {author} {\bibinfo {author} {\bibfnamefont {D.~P.}\ \bibnamefont
  {Norton}}, \bibinfo {author} {\bibfnamefont {N.~A.}\ \bibnamefont
  {Theodoropoulou}}, \bibinfo {author} {\bibfnamefont {A.~F.}\ \bibnamefont
  {Hebard}}, \bibinfo {author} {\bibfnamefont {J.~D.}\ \bibnamefont {Budai}},
  \bibinfo {author} {\bibfnamefont {L.~A.}\ \bibnamefont {Boatner}}, \bibinfo
  {author} {\bibfnamefont {S.~J.}\ \bibnamefont {Pearton}}, \ and\ \bibinfo
  {author} {\bibfnamefont {R.~G.}\ \bibnamefont {Wilson}},\ }\href {\doibase
  10.1149/1.1531871} {\bibfield  {journal} {\bibinfo  {journal}
  {Electrochemical and Solid-State Letters,}\ }\textbf {\bibinfo {volume}
  {6}},\ \bibinfo {pages} {G19} (\bibinfo {year} {2003})}\BibitemShut {NoStop}%
\bibitem [{\citenamefont {Chandra}\ \emph {et~al.}(2013)\citenamefont
  {Chandra}, \citenamefont {Gupta}, \citenamefont {Nandy},\ and\ \citenamefont
  {Mahadevan}}]{PhysRevB.87.214110}%
  \BibitemOpen
  \bibfield  {author} {\bibinfo {author} {\bibfnamefont {H.~K.}\ \bibnamefont
  {Chandra}}, \bibinfo {author} {\bibfnamefont {K.}~\bibnamefont {Gupta}},
  \bibinfo {author} {\bibfnamefont {A.~K.}\ \bibnamefont {Nandy}}, \ and\
  \bibinfo {author} {\bibfnamefont {P.}~\bibnamefont {Mahadevan}},\ }\href
  {\doibase 10.1103/PhysRevB.87.214110} {\bibfield  {journal} {\bibinfo
  {journal} {Phys. Rev. B}\ }\textbf {\bibinfo {volume} {87}},\ \bibinfo
  {pages} {214110} (\bibinfo {year} {2013})}\BibitemShut {NoStop}%
\bibitem [{\citenamefont {Shuai}\ \emph {et~al.}(2011)\citenamefont {Shuai},
  \citenamefont {Zhou}, \citenamefont {Bürger}, \citenamefont {Reuther},
  \citenamefont {Skorupa}, \citenamefont {John}, \citenamefont {Helm},\ and\
  \citenamefont {Schmidt}}]{JAP1}%
  \BibitemOpen
  \bibfield  {author} {\bibinfo {author} {\bibfnamefont {Y.}~\bibnamefont
  {Shuai}}, \bibinfo {author} {\bibfnamefont {S.}~\bibnamefont {Zhou}},
  \bibinfo {author} {\bibfnamefont {D.}~\bibnamefont {Bürger}}, \bibinfo
  {author} {\bibfnamefont {H.}~\bibnamefont {Reuther}}, \bibinfo {author}
  {\bibfnamefont {I.}~\bibnamefont {Skorupa}}, \bibinfo {author} {\bibfnamefont
  {V.}~\bibnamefont {John}}, \bibinfo {author} {\bibfnamefont {M.}~\bibnamefont
  {Helm}}, \ and\ \bibinfo {author} {\bibfnamefont {H.}~\bibnamefont
  {Schmidt}},\ }\href@noop {} {\bibfield  {journal} {\bibinfo  {journal} {J.
  Appl. Phys.}\ }\textbf {\bibinfo {volume} {109}},\ \bibinfo {eid} {084105}
  (\bibinfo {year} {2011})}\BibitemShut {NoStop}%
\bibitem [{\citenamefont {Tong}\ \emph {et~al.}(2008)\citenamefont {Tong},
  \citenamefont {Lin}, \citenamefont {Zhang}, \citenamefont {Wang},\ and\
  \citenamefont {Nan}}]{JAP2}%
  \BibitemOpen
  \bibfield  {author} {\bibinfo {author} {\bibfnamefont {X.}~\bibnamefont
  {Tong}}, \bibinfo {author} {\bibfnamefont {Y.-H.}\ \bibnamefont {Lin}},
  \bibinfo {author} {\bibfnamefont {S.}~\bibnamefont {Zhang}}, \bibinfo
  {author} {\bibfnamefont {Y.}~\bibnamefont {Wang}}, \ and\ \bibinfo {author}
  {\bibfnamefont {C.-W.}\ \bibnamefont {Nan}},\ }\href {\doibase
  http://dx.doi.org/10.1063/1.2973202} {\bibfield  {journal} {\bibinfo
  {journal} {Journal of Applied Physics}\ }\textbf {\bibinfo {volume} {104}},\
  \bibinfo {eid} {066108} (\bibinfo {year} {2008})}\BibitemShut {NoStop}%
\bibitem [{\citenamefont {Bottcher}\ \emph {et~al.}(2005)\citenamefont
  {Bottcher}, \citenamefont {Langhammer}, \citenamefont {Muller},\ and\
  \citenamefont {Abicht}}]{aip1}%
  \BibitemOpen
  \bibfield  {author} {\bibinfo {author} {\bibfnamefont {R.}~\bibnamefont
  {Bottcher}}, \bibinfo {author} {\bibfnamefont {H.~T.}\ \bibnamefont
  {Langhammer}}, \bibinfo {author} {\bibfnamefont {T.}~\bibnamefont {Muller}},
  \ and\ \bibinfo {author} {\bibfnamefont {H.-P.}\ \bibnamefont {Abicht}},\
  }\href@noop {} {\bibfield  {journal} {\bibinfo  {journal} {J. Phys.: Condens.
  Matt.}\ }\textbf {\bibinfo {volume} {17}},\ \bibinfo {pages} {4925} (\bibinfo
  {year} {2005})}\BibitemShut {NoStop}%
\bibitem [{\citenamefont {Jayakumar}\ \emph {et~al.}(2006)\citenamefont
  {Jayakumar}, \citenamefont {Salunke}, \citenamefont {Kadam}, \citenamefont
  {Mohapatra}, \citenamefont {Yaswant},\ and\ \citenamefont
  {Kulshreshtha}}]{0957-4484-17-5-020}%
  \BibitemOpen
  \bibfield  {author} {\bibinfo {author} {\bibfnamefont {O.~D.}\ \bibnamefont
  {Jayakumar}}, \bibinfo {author} {\bibfnamefont {H.~G.}\ \bibnamefont
  {Salunke}}, \bibinfo {author} {\bibfnamefont {R.~M.}\ \bibnamefont {Kadam}},
  \bibinfo {author} {\bibfnamefont {M.}~\bibnamefont {Mohapatra}}, \bibinfo
  {author} {\bibfnamefont {G.}~\bibnamefont {Yaswant}}, \ and\ \bibinfo
  {author} {\bibfnamefont {S.~K.}\ \bibnamefont {Kulshreshtha}},\ }\href@noop
  {} {\bibfield  {journal} {\bibinfo  {journal} {Nanotechnology}\ }\textbf
  {\bibinfo {volume} {17}},\ \bibinfo {pages} {1278} (\bibinfo {year}
  {2006})}\BibitemShut {NoStop}%
\bibitem [{\citenamefont {Bottcher}\ \emph {et~al.}(2008)\citenamefont
  {Bottcher}, \citenamefont {Langhammer}, \citenamefont {Muller},\ and\
  \citenamefont {Abicht}}]{0953-8984-20-50-505209}%
  \BibitemOpen
  \bibfield  {author} {\bibinfo {author} {\bibfnamefont {R.}~\bibnamefont
  {Bottcher}}, \bibinfo {author} {\bibfnamefont {H.~T.}\ \bibnamefont
  {Langhammer}}, \bibinfo {author} {\bibfnamefont {T.}~\bibnamefont {Muller}},
  \ and\ \bibinfo {author} {\bibfnamefont {H.-P.}\ \bibnamefont {Abicht}},\
  }\href {http://stacks.iop.org/0953-8984/20/i=50/a=505209} {\bibfield
  {journal} {\bibinfo  {journal} {J. Phys. Cond. Matt.}\ }\textbf {\bibinfo
  {volume} {20}},\ \bibinfo {pages} {505209} (\bibinfo {year}
  {2008})}\BibitemShut {NoStop}%
\bibitem [{\citenamefont {Langhammer}\ \emph {et~al.}(2008)\citenamefont
  {Langhammer}, \citenamefont {Muller}, \citenamefont {Böttcher},\ and\
  \citenamefont {Abicht}}]{0953-8984-20-8-085206}%
  \BibitemOpen
  \bibfield  {author} {\bibinfo {author} {\bibfnamefont {H.~T.}\ \bibnamefont
  {Langhammer}}, \bibinfo {author} {\bibfnamefont {T.}~\bibnamefont {Muller}},
  \bibinfo {author} {\bibfnamefont {R.}~\bibnamefont {Böttcher}}, \ and\
  \bibinfo {author} {\bibfnamefont {H.-P.}\ \bibnamefont {Abicht}},\ }\href
  {http://stacks.iop.org/0953-8984/20/i=8/a=085206} {\bibfield  {journal}
  {\bibinfo  {journal} {J. Phys. Cond. Matt.}\ }\textbf {\bibinfo {volume}
  {20}},\ \bibinfo {pages} {085206} (\bibinfo {year} {2008})}\BibitemShut
  {NoStop}%
\bibitem [{\citenamefont {Wu}\ \emph {et~al.}(2009)\citenamefont {Wu},
  \citenamefont {Fang}, \citenamefont {Feng},\ and\ \citenamefont
  {Zheng}}]{doi:742066}%
  \BibitemOpen
  \bibfield  {author} {\bibinfo {author} {\bibfnamefont {X.-X.}\ \bibnamefont
  {Wu}}, \bibinfo {author} {\bibfnamefont {W.}~\bibnamefont {Fang}}, \bibinfo
  {author} {\bibfnamefont {W.-L.}\ \bibnamefont {Feng}}, \ and\ \bibinfo
  {author} {\bibfnamefont {W.-C.}\ \bibnamefont {Zheng}},\ }\href@noop {}
  {\bibfield  {journal} {\bibinfo  {journal} {Pramana}\ }\textbf {\bibinfo
  {volume} {72}},\ \bibinfo {pages} {569} (\bibinfo {year} {2009})}\BibitemShut
  {NoStop}%
\bibitem [{\citenamefont {Dang}\ \emph {et~al.}(2011)\citenamefont {Dang},
  \citenamefont {Thanh}, \citenamefont {Hong}, \citenamefont {Lam},\ and\
  \citenamefont {Phan}}]{Dang}%
  \BibitemOpen
  \bibfield  {author} {\bibinfo {author} {\bibfnamefont {N.~V.}\ \bibnamefont
  {Dang}}, \bibinfo {author} {\bibfnamefont {T.~D.}\ \bibnamefont {Thanh}},
  \bibinfo {author} {\bibfnamefont {L.~V.}\ \bibnamefont {Hong}}, \bibinfo
  {author} {\bibfnamefont {V.~D.}\ \bibnamefont {Lam}}, \ and\ \bibinfo
  {author} {\bibfnamefont {T.-L.}\ \bibnamefont {Phan}},\ }\href {\doibase
  http://dx.doi.org/10.1063/1.3625235} {\bibfield  {journal} {\bibinfo
  {journal} {J. App. Phys.}\ }\textbf {\bibinfo {volume} {110}},\ \bibinfo
  {eid} {043914} (\bibinfo {year} {2011})}\BibitemShut {NoStop}%
\bibitem [{\citenamefont {Dang}\ \emph {et~al.}(2012)\citenamefont {Dang},
  \citenamefont {Phan}, \citenamefont {Thanh}, \citenamefont {Lam},\ and\
  \citenamefont {Hong}}]{doi:1499359}%
  \BibitemOpen
  \bibfield  {author} {\bibinfo {author} {\bibfnamefont {N.~V.}\ \bibnamefont
  {Dang}}, \bibinfo {author} {\bibfnamefont {T.-L.}\ \bibnamefont {Phan}},
  \bibinfo {author} {\bibfnamefont {T.~D.}\ \bibnamefont {Thanh}}, \bibinfo
  {author} {\bibfnamefont {V.~D.}\ \bibnamefont {Lam}}, \ and\ \bibinfo
  {author} {\bibfnamefont {L.~V.}\ \bibnamefont {Hong}},\ }\href@noop {}
  {\bibfield  {journal} {\bibinfo  {journal} {Journal of Applied Physics}\
  }\textbf {\bibinfo {volume} {111}},\ \bibinfo {eid} {113913} (\bibinfo {year}
  {2012})}\BibitemShut {NoStop}%
\bibitem [{\citenamefont {Phan}\ \emph {et~al.}(2012)\citenamefont {Phan},
  \citenamefont {Zhang}, \citenamefont {Grinting}, \citenamefont {Yu},
  \citenamefont {Nghia}, \citenamefont {Dang},\ and\ \citenamefont
  {Lam}}]{doi:1511757}%
  \BibitemOpen
  \bibfield  {author} {\bibinfo {author} {\bibfnamefont {T.-L.}\ \bibnamefont
  {Phan}}, \bibinfo {author} {\bibfnamefont {P.}~\bibnamefont {Zhang}},
  \bibinfo {author} {\bibfnamefont {D.}~\bibnamefont {Grinting}}, \bibinfo
  {author} {\bibfnamefont {S.~C.}\ \bibnamefont {Yu}}, \bibinfo {author}
  {\bibfnamefont {N.~X.}\ \bibnamefont {Nghia}}, \bibinfo {author}
  {\bibfnamefont {N.~V.}\ \bibnamefont {Dang}}, \ and\ \bibinfo {author}
  {\bibfnamefont {V.~D.}\ \bibnamefont {Lam}},\ }\href@noop {} {\bibfield
  {journal} {\bibinfo  {journal} {Journal of Applied Physics}\ }\textbf
  {\bibinfo {volume} {112}},\ \bibinfo {eid} {013909} (\bibinfo {year}
  {2012})}\BibitemShut {NoStop}%
\bibitem [{\citenamefont {Gupta}\ \emph {et~al.}(2014)\citenamefont {Gupta},
  \citenamefont {Kadam}, \citenamefont {Gupta}, \citenamefont {Sahu},\ and\
  \citenamefont {Natarajan}}]{doi:1812576}%
  \BibitemOpen
  \bibfield  {author} {\bibinfo {author} {\bibfnamefont {S.~K.}\ \bibnamefont
  {Gupta}}, \bibinfo {author} {\bibfnamefont {R.}~\bibnamefont {Kadam}},
  \bibinfo {author} {\bibfnamefont {R.}~\bibnamefont {Gupta}}, \bibinfo
  {author} {\bibfnamefont {M.}~\bibnamefont {Sahu}}, \ and\ \bibinfo {author}
  {\bibfnamefont {V.}~\bibnamefont {Natarajan}},\ }\href {\doibase
  http://dx.doi.org/10.1016/j.matchemphys.2014.01.054} {\bibfield  {journal}
  {\bibinfo  {journal} {Mat. Chem. Phys.}\ }\textbf {\bibinfo {volume} {145}},\
  \bibinfo {pages} {162 } (\bibinfo {year} {2014})}\BibitemShut {NoStop}%
\bibitem [{\citenamefont {Ren}(2004)}]{Ren2004}%
  \BibitemOpen
  \bibfield  {author} {\bibinfo {author} {\bibfnamefont {X.}~\bibnamefont
  {Ren}},\ }\href {\doibase 10.1038/nmat1051} {\bibfield  {journal} {\bibinfo
  {journal} {Nature Materials}\ }\textbf {\bibinfo {volume} {3}},\ \bibinfo
  {pages} {91} (\bibinfo {year} {2004})}\BibitemShut {NoStop}%
\bibitem [{\citenamefont {Zhang}\ and\ \citenamefont {Ren}(2005)}]{Zhang2005}%
  \BibitemOpen
  \bibfield  {author} {\bibinfo {author} {\bibfnamefont {L.~X.}\ \bibnamefont
  {Zhang}}\ and\ \bibinfo {author} {\bibfnamefont {X.}~\bibnamefont {Ren}},\
  }\href {<Go to ISI>://WOS:000229935000028} {\bibfield  {journal} {\bibinfo
  {journal} {Physical Review B}\ }\textbf {\bibinfo {volume} {71}} (\bibinfo
  {year} {2005})}\BibitemShut {NoStop}%
\bibitem [{\citenamefont {Nossa}\ \emph {et~al.}(2015)\citenamefont {Nossa},
  \citenamefont {Naumov},\ and\ \citenamefont {Cohen}}]{Nossa}%
  \BibitemOpen
  \bibfield  {author} {\bibinfo {author} {\bibfnamefont {J.~F.}\ \bibnamefont
  {Nossa}}, \bibinfo {author} {\bibfnamefont {I.~I.}\ \bibnamefont {Naumov}}, \
  and\ \bibinfo {author} {\bibfnamefont {R.~E.}\ \bibnamefont {Cohen}},\ }\href
  {\doibase 10.1103/PhysRevB.91.214105} {\bibfield  {journal} {\bibinfo
  {journal} {Phys. Rev. B}\ }\textbf {\bibinfo {volume} {91}},\ \bibinfo
  {pages} {214105} (\bibinfo {year} {2015})}\BibitemShut {NoStop}%
\bibitem [{\citenamefont {Chapman}\ \emph {et~al.}(2017)\citenamefont
  {Chapman}, \citenamefont {Cohen}, \citenamefont {Kimmel},\ and\ \citenamefont
  {Duffy}}]{Chapman2017}%
  \BibitemOpen
  \bibfield  {author} {\bibinfo {author} {\bibfnamefont {J.~B.~J.}\
  \bibnamefont {Chapman}}, \bibinfo {author} {\bibfnamefont {R.~E.}\
  \bibnamefont {Cohen}}, \bibinfo {author} {\bibfnamefont {A.~V.}\ \bibnamefont
  {Kimmel}}, \ and\ \bibinfo {author} {\bibfnamefont {D.~M.}\ \bibnamefont
  {Duffy}},\ }\href {\doibase 10.1103/PhysRevLett.119.177602} {\bibfield
  {journal} {\bibinfo  {journal} {Phys Rev Lett}\ }\textbf {\bibinfo {volume}
  {119}},\ \bibinfo {pages} {177602} (\bibinfo {year} {2017})}\BibitemShut
  {NoStop}%
\bibitem [{\citenamefont {Liu}\ and\ \citenamefont {Cohen}(2017)}]{Liu2017}%
  \BibitemOpen
  \bibfield  {author} {\bibinfo {author} {\bibfnamefont {S.}~\bibnamefont
  {Liu}}\ and\ \bibinfo {author} {\bibfnamefont {R.~E.}\ \bibnamefont
  {Cohen}},\ }\href {\doibase 10.1063/1.4989670} {\bibfield  {journal}
  {\bibinfo  {journal} {Applied Physics Letters}\ }\textbf {\bibinfo {volume}
  {111}},\ \bibinfo {pages} {082903} (\bibinfo {year} {2017})}\BibitemShut
  {NoStop}%
\bibitem [{\citenamefont {Dietl}(2010)}]{RN14480}%
  \BibitemOpen
  \bibfield  {author} {\bibinfo {author} {\bibfnamefont {T.}~\bibnamefont
  {Dietl}},\ }\href {\doibase 10.1038/nmat2898} {\bibfield  {journal} {\bibinfo
   {journal} {Nat Mater}\ }\textbf {\bibinfo {volume} {9}},\ \bibinfo {pages}
  {965} (\bibinfo {year} {2010})}\BibitemShut {NoStop}%
\bibitem [{\citenamefont {Sato}\ \emph {et~al.}(2010)\citenamefont {Sato},
  \citenamefont {Bergqvist}, \citenamefont {KudrnovskÃœ}, \citenamefont
  {Dederichs}, \citenamefont {Eriksson}, \citenamefont {Turek}, \citenamefont
  {Sanyal}, \citenamefont {Bouzerar}, \citenamefont {Katayama-Yoshida},
  \citenamefont {Dinh}, \citenamefont {Fukushima}, \citenamefont {Kizaki},\
  and\ \citenamefont {Zeller}}]{RN14521}%
  \BibitemOpen
  \bibfield  {author} {\bibinfo {author} {\bibfnamefont {K.}~\bibnamefont
  {Sato}}, \bibinfo {author} {\bibfnamefont {L.}~\bibnamefont {Bergqvist}},
  \bibinfo {author} {\bibfnamefont {J.}~\bibnamefont {KudrnovskÃœ}}, \bibinfo
  {author} {\bibfnamefont {P.~H.}\ \bibnamefont {Dederichs}}, \bibinfo {author}
  {\bibfnamefont {O.}~\bibnamefont {Eriksson}}, \bibinfo {author}
  {\bibfnamefont {I.}~\bibnamefont {Turek}}, \bibinfo {author} {\bibfnamefont
  {B.}~\bibnamefont {Sanyal}}, \bibinfo {author} {\bibfnamefont
  {G.}~\bibnamefont {Bouzerar}}, \bibinfo {author} {\bibfnamefont
  {H.}~\bibnamefont {Katayama-Yoshida}}, \bibinfo {author} {\bibfnamefont
  {V.~A.}\ \bibnamefont {Dinh}}, \bibinfo {author} {\bibfnamefont
  {T.}~\bibnamefont {Fukushima}}, \bibinfo {author} {\bibfnamefont
  {H.}~\bibnamefont {Kizaki}}, \ and\ \bibinfo {author} {\bibfnamefont
  {R.}~\bibnamefont {Zeller}},\ }\href {\doibase 10.1103/RevModPhys.82.1633}
  {\bibfield  {journal} {\bibinfo  {journal} {Reviews of Modern Physics}\
  }\textbf {\bibinfo {volume} {82}},\ \bibinfo {pages} {1633} (\bibinfo {year}
  {2010})}\BibitemShut {NoStop}%
\bibitem [{\citenamefont {Pauling}(1960)}]{pauling}%
  \BibitemOpen
  \bibfield  {author} {\bibinfo {author} {\bibfnamefont {L.}~\bibnamefont
  {Pauling}},\ }\href@noop {} {\emph {\bibinfo {title} {The nature of the
  chemical bond}}},\ \bibinfo {edition} {3rd}\ ed.\ (\bibinfo  {publisher}
  {Cornell University Press},\ \bibinfo {year} {1960})\BibitemShut {NoStop}%
\bibitem [{\citenamefont {Quan}\ \emph {et~al.}(2012)\citenamefont {Quan},
  \citenamefont {Pardo},\ and\ \citenamefont
  {Pickett}}]{PhysRevLett.109.216401}%
  \BibitemOpen
  \bibfield  {author} {\bibinfo {author} {\bibfnamefont {Y.}~\bibnamefont
  {Quan}}, \bibinfo {author} {\bibfnamefont {V.}~\bibnamefont {Pardo}}, \ and\
  \bibinfo {author} {\bibfnamefont {W.~E.}\ \bibnamefont {Pickett}},\ }\href
  {\doibase 10.1103/PhysRevLett.109.216401} {\bibfield  {journal} {\bibinfo
  {journal} {Phys. Rev. Lett.}\ }\textbf {\bibinfo {volume} {109}},\ \bibinfo
  {pages} {216401} (\bibinfo {year} {2012})}\BibitemShut {NoStop}%
\bibitem [{\citenamefont {Jiang}\ \emph {et~al.}(2012)\citenamefont {Jiang},
  \citenamefont {Levchenko},\ and\ \citenamefont
  {Rappe}}]{PhysRevLett.108.166403}%
  \BibitemOpen
  \bibfield  {author} {\bibinfo {author} {\bibfnamefont {L.}~\bibnamefont
  {Jiang}}, \bibinfo {author} {\bibfnamefont {S.~V.}\ \bibnamefont
  {Levchenko}}, \ and\ \bibinfo {author} {\bibfnamefont {A.~M.}\ \bibnamefont
  {Rappe}},\ }\href {\doibase 10.1103/PhysRevLett.108.166403} {\bibfield
  {journal} {\bibinfo  {journal} {Phys. Rev. Lett.}\ }\textbf {\bibinfo
  {volume} {108}},\ \bibinfo {pages} {166403} (\bibinfo {year}
  {2012})}\BibitemShut {NoStop}%
\bibitem [{\citenamefont {Sit}\ \emph {et~al.}(2011)\citenamefont {Sit},
  \citenamefont {Car}, \citenamefont {Cohen},\ and\ \citenamefont
  {Selloni}}]{car1}%
  \BibitemOpen
  \bibfield  {author} {\bibinfo {author} {\bibfnamefont {P.~H.-L.}\
  \bibnamefont {Sit}}, \bibinfo {author} {\bibfnamefont {R.}~\bibnamefont
  {Car}}, \bibinfo {author} {\bibfnamefont {M.~H.}\ \bibnamefont {Cohen}}, \
  and\ \bibinfo {author} {\bibfnamefont {A.}~\bibnamefont {Selloni}},\ }\href
  {\doibase 10.1021/ic2013107} {\bibfield  {journal} {\bibinfo  {journal}
  {Inorganic Chemistry}\ }\textbf {\bibinfo {volume} {50}},\ \bibinfo {pages}
  {10259} (\bibinfo {year} {2011})}\BibitemShut {NoStop}%
\bibitem [{\citenamefont {Pickett}\ \emph {et~al.}(2014)\citenamefont
  {Pickett}, \citenamefont {Quan},\ and\ \citenamefont {Pardo}}]{pk2}%
  \BibitemOpen
  \bibfield  {author} {\bibinfo {author} {\bibfnamefont {W.~E.}\ \bibnamefont
  {Pickett}}, \bibinfo {author} {\bibfnamefont {Y.}~\bibnamefont {Quan}}, \
  and\ \bibinfo {author} {\bibfnamefont {V.}~\bibnamefont {Pardo}},\ }\href
  {http://stacks.iop.org/0953-8984/26/i=27/a=274203} {\bibfield  {journal}
  {\bibinfo  {journal} {Journal of Physics: Condensed Matter}\ }\textbf
  {\bibinfo {volume} {26}},\ \bibinfo {pages} {274203} (\bibinfo {year}
  {2014})}\BibitemShut {NoStop}%
\bibitem [{\citenamefont {Cheng}\ \emph {et~al.}(2012)\citenamefont {Cheng},
  \citenamefont {Hu}, \citenamefont {Chen}, \citenamefont {Xu}, \citenamefont
  {Zheng}, \citenamefont {Luo},\ and\ \citenamefont
  {Wang}}]{PhysRevB.86.134503}%
  \BibitemOpen
  \bibfield  {author} {\bibinfo {author} {\bibfnamefont {B.}~\bibnamefont
  {Cheng}}, \bibinfo {author} {\bibfnamefont {B.~F.}\ \bibnamefont {Hu}},
  \bibinfo {author} {\bibfnamefont {R.~Y.}\ \bibnamefont {Chen}}, \bibinfo
  {author} {\bibfnamefont {G.}~\bibnamefont {Xu}}, \bibinfo {author}
  {\bibfnamefont {P.}~\bibnamefont {Zheng}}, \bibinfo {author} {\bibfnamefont
  {J.~L.}\ \bibnamefont {Luo}}, \ and\ \bibinfo {author} {\bibfnamefont
  {N.~L.}\ \bibnamefont {Wang}},\ }\href {\doibase 10.1103/PhysRevB.86.134503}
  {\bibfield  {journal} {\bibinfo  {journal} {Phys. Rev. B}\ }\textbf {\bibinfo
  {volume} {86}},\ \bibinfo {pages} {134503} (\bibinfo {year}
  {2012})}\BibitemShut {NoStop}%
\bibitem [{\citenamefont {Georgescu}\ and\ \citenamefont
  {Ismail-Beigi}(2017)}]{PhysRevB.96.165135}%
  \BibitemOpen
  \bibfield  {author} {\bibinfo {author} {\bibfnamefont {A.~B.}\ \bibnamefont
  {Georgescu}}\ and\ \bibinfo {author} {\bibfnamefont {S.}~\bibnamefont
  {Ismail-Beigi}},\ }\href {\doibase 10.1103/PhysRevB.96.165135} {\bibfield
  {journal} {\bibinfo  {journal} {Phys. Rev. B}\ }\textbf {\bibinfo {volume}
  {96}},\ \bibinfo {pages} {165135} (\bibinfo {year} {2017})}\BibitemShut
  {NoStop}%
\bibitem [{\citenamefont {Nakayama}\ and\ \citenamefont
  {Katayama-Yoshida}(2001)}]{lda1}%
  \BibitemOpen
  \bibfield  {author} {\bibinfo {author} {\bibfnamefont {H.}~\bibnamefont
  {Nakayama}}\ and\ \bibinfo {author} {\bibfnamefont {H.}~\bibnamefont
  {Katayama-Yoshida}},\ }\href@noop {} {\bibfield  {journal} {\bibinfo
  {journal} {Jpn. J. Appl. Phys.}\ }\textbf {\bibinfo {volume} {40}},\ \bibinfo
  {pages} {L1355} (\bibinfo {year} {2001})}\BibitemShut {NoStop}%
\bibitem [{\citenamefont {Kotliar}\ \emph {et~al.}(2006)\citenamefont
  {Kotliar}, \citenamefont {Savrasov}, \citenamefont {Haule}, \citenamefont
  {Oudovenko}, \citenamefont {Parcollet},\ and\ \citenamefont
  {Marianetti}}]{RevModPhys.78.865}%
  \BibitemOpen
  \bibfield  {author} {\bibinfo {author} {\bibfnamefont {G.}~\bibnamefont
  {Kotliar}}, \bibinfo {author} {\bibfnamefont {S.~Y.}\ \bibnamefont
  {Savrasov}}, \bibinfo {author} {\bibfnamefont {K.}~\bibnamefont {Haule}},
  \bibinfo {author} {\bibfnamefont {V.~S.}\ \bibnamefont {Oudovenko}}, \bibinfo
  {author} {\bibfnamefont {O.}~\bibnamefont {Parcollet}}, \ and\ \bibinfo
  {author} {\bibfnamefont {C.~A.}\ \bibnamefont {Marianetti}},\ }\href
  {\doibase 10.1103/RevModPhys.78.865} {\bibfield  {journal} {\bibinfo
  {journal} {Rev. Mod. Phys.}\ }\textbf {\bibinfo {volume} {78}},\ \bibinfo
  {pages} {865} (\bibinfo {year} {2006})}\BibitemShut {NoStop}%
\bibitem [{\citenamefont {Shim}\ \emph {et~al.}(2007)\citenamefont {Shim},
  \citenamefont {Haule},\ and\ \citenamefont {Kotliar}}]{Shim}%
  \BibitemOpen
  \bibfield  {author} {\bibinfo {author} {\bibfnamefont {J.~H.}\ \bibnamefont
  {Shim}}, \bibinfo {author} {\bibfnamefont {K.}~\bibnamefont {Haule}}, \ and\
  \bibinfo {author} {\bibfnamefont {G.}~\bibnamefont {Kotliar}},\ }\href
  {http://dx.doi.org/10.1038/nature05647} {\bibfield  {journal} {\bibinfo
  {journal} {Nature}\ }\textbf {\bibinfo {volume} {446}},\ \bibinfo {pages}
  {513} (\bibinfo {year} {2007})}\BibitemShut {NoStop}%
\bibitem [{\citenamefont {Zhang}\ \emph {et~al.}(2015)\citenamefont {Zhang},
  \citenamefont {Cohen},\ and\ \citenamefont {Haule}}]{peng}%
  \BibitemOpen
  \bibfield  {author} {\bibinfo {author} {\bibfnamefont {P.}~\bibnamefont
  {Zhang}}, \bibinfo {author} {\bibfnamefont {R.~E.}\ \bibnamefont {Cohen}}, \
  and\ \bibinfo {author} {\bibfnamefont {K.}~\bibnamefont {Haule}},\
  }\href@noop {} {\bibfield  {journal} {\bibinfo  {journal} {Nature}\ }\textbf
  {\bibinfo {volume} {517}},\ \bibinfo {pages} {605} (\bibinfo {year}
  {2015})}\BibitemShut {NoStop}%
\bibitem [{\citenamefont {Yin}\ \emph {et~al.}(2011)\citenamefont {Yin},
  \citenamefont {Haule},\ and\ \citenamefont {Kotliar}}]{haule3}%
  \BibitemOpen
  \bibfield  {author} {\bibinfo {author} {\bibfnamefont {Z.~P.}\ \bibnamefont
  {Yin}}, \bibinfo {author} {\bibfnamefont {K.}~\bibnamefont {Haule}}, \ and\
  \bibinfo {author} {\bibfnamefont {G.}~\bibnamefont {Kotliar}},\ }\href@noop
  {} {\bibfield  {journal} {\bibinfo  {journal} {Nat. Mater.}\ }\textbf
  {\bibinfo {volume} {10}},\ \bibinfo {pages} {932} (\bibinfo {year}
  {2011})}\BibitemShut {NoStop}%
\bibitem [{\citenamefont {Liu}\ \emph {et~al.}(2012)\citenamefont {Liu},
  \citenamefont {Harriger}, \citenamefont {Luo}, \citenamefont {Wang},
  \citenamefont {Ewings}, \citenamefont {Guidi}, \citenamefont {Park},
  \citenamefont {Haule}, \citenamefont {Kotliar}, \citenamefont {Hayden},\ and\
  \citenamefont {Dai}}]{haule_spin}%
  \BibitemOpen
  \bibfield  {author} {\bibinfo {author} {\bibfnamefont {M.}~\bibnamefont
  {Liu}}, \bibinfo {author} {\bibfnamefont {L.~W.}\ \bibnamefont {Harriger}},
  \bibinfo {author} {\bibfnamefont {H.}~\bibnamefont {Luo}}, \bibinfo {author}
  {\bibfnamefont {M.}~\bibnamefont {Wang}}, \bibinfo {author} {\bibfnamefont
  {R.~A.}\ \bibnamefont {Ewings}}, \bibinfo {author} {\bibfnamefont
  {T.}~\bibnamefont {Guidi}}, \bibinfo {author} {\bibfnamefont
  {H.}~\bibnamefont {Park}}, \bibinfo {author} {\bibfnamefont {K.}~\bibnamefont
  {Haule}}, \bibinfo {author} {\bibfnamefont {G.}~\bibnamefont {Kotliar}},
  \bibinfo {author} {\bibfnamefont {S.~M.}\ \bibnamefont {Hayden}}, \ and\
  \bibinfo {author} {\bibfnamefont {P.}~\bibnamefont {Dai}},\ }\href@noop {}
  {\bibfield  {journal} {\bibinfo  {journal} {Nat. Phys.}\ }\textbf {\bibinfo
  {volume} {8}},\ \bibinfo {pages} {376} (\bibinfo {year} {2012})}\BibitemShut
  {NoStop}%
\bibitem [{\citenamefont {Wang}\ \emph {et~al.}(2013)\citenamefont {Wang},
  \citenamefont {Zhang}, \citenamefont {Lu}, \citenamefont {Tan}, \citenamefont
  {Luo}, \citenamefont {Song}, \citenamefont {Wang}, \citenamefont {Zhang},
  \citenamefont {Goremychkin}, \citenamefont {Perring}, \citenamefont {Maier},
  \citenamefont {Yin}, \citenamefont {Haule}, \citenamefont {Kotliar},\ and\
  \citenamefont {Dai}}]{Wang:2013gz}%
  \BibitemOpen
  \bibfield  {author} {\bibinfo {author} {\bibfnamefont {M.}~\bibnamefont
  {Wang}}, \bibinfo {author} {\bibfnamefont {C.}~\bibnamefont {Zhang}},
  \bibinfo {author} {\bibfnamefont {X.}~\bibnamefont {Lu}}, \bibinfo {author}
  {\bibfnamefont {G.}~\bibnamefont {Tan}}, \bibinfo {author} {\bibfnamefont
  {H.}~\bibnamefont {Luo}}, \bibinfo {author} {\bibfnamefont {Y.}~\bibnamefont
  {Song}}, \bibinfo {author} {\bibfnamefont {M.}~\bibnamefont {Wang}}, \bibinfo
  {author} {\bibfnamefont {X.}~\bibnamefont {Zhang}}, \bibinfo {author}
  {\bibfnamefont {E.~A.}\ \bibnamefont {Goremychkin}}, \bibinfo {author}
  {\bibfnamefont {T.~G.}\ \bibnamefont {Perring}}, \bibinfo {author}
  {\bibfnamefont {T.~A.}\ \bibnamefont {Maier}}, \bibinfo {author}
  {\bibfnamefont {Z.}~\bibnamefont {Yin}}, \bibinfo {author} {\bibfnamefont
  {K.}~\bibnamefont {Haule}}, \bibinfo {author} {\bibfnamefont
  {G.}~\bibnamefont {Kotliar}}, \ and\ \bibinfo {author} {\bibfnamefont
  {P.}~\bibnamefont {Dai}},\ }\href@noop {} {\bibfield  {journal} {\bibinfo
  {journal} {Nat. Comm.}\ }\textbf {\bibinfo {volume} {4}},\ \bibinfo {pages}
  {1} (\bibinfo {year} {2013})}\BibitemShut {NoStop}%
\bibitem [{\citenamefont {Schafgans}\ \emph {et~al.}(2012)\citenamefont
  {Schafgans}, \citenamefont {Pursley}, \citenamefont {LaForge}, \citenamefont
  {Qazilbash}, \citenamefont {Sefat}, \citenamefont {Mandrus}, \citenamefont
  {Haule}, \citenamefont {Kotliar},\ and\ \citenamefont {Basov}}]{Schafgans}%
  \BibitemOpen
  \bibfield  {author} {\bibinfo {author} {\bibfnamefont {S.~J.}\ \bibnamefont
  {Schafgans}, \bibfnamefont {A.~A .and~Moon}}, \bibinfo {author}
  {\bibfnamefont {B.~C.}\ \bibnamefont {Pursley}}, \bibinfo {author}
  {\bibfnamefont {A.~D.}\ \bibnamefont {LaForge}}, \bibinfo {author}
  {\bibfnamefont {M.~M.}\ \bibnamefont {Qazilbash}}, \bibinfo {author}
  {\bibfnamefont {A.~S.}\ \bibnamefont {Sefat}}, \bibinfo {author}
  {\bibfnamefont {D.}~\bibnamefont {Mandrus}}, \bibinfo {author} {\bibfnamefont
  {K.}~\bibnamefont {Haule}}, \bibinfo {author} {\bibfnamefont
  {G.}~\bibnamefont {Kotliar}}, \ and\ \bibinfo {author} {\bibfnamefont
  {D.~N.}\ \bibnamefont {Basov}},\ }\href@noop {} {\bibfield  {journal}
  {\bibinfo  {journal} {Phys. Rev. Lett.}\ }\textbf {\bibinfo {volume} {108}},\
  \bibinfo {pages} {147002} (\bibinfo {year} {2012})}\BibitemShut {NoStop}%
\bibitem [{\citenamefont {Mandal}\ \emph
  {et~al.}(2014{\natexlab{a}})\citenamefont {Mandal}, \citenamefont {Cohen},\
  and\ \citenamefont {Haule}}]{PhysRevB.89.220502}%
  \BibitemOpen
  \bibfield  {author} {\bibinfo {author} {\bibfnamefont {S.}~\bibnamefont
  {Mandal}}, \bibinfo {author} {\bibfnamefont {R.~E.}\ \bibnamefont {Cohen}}, \
  and\ \bibinfo {author} {\bibfnamefont {K.}~\bibnamefont {Haule}},\ }\href
  {\doibase 10.1103/PhysRevB.89.220502} {\bibfield  {journal} {\bibinfo
  {journal} {Phys. Rev. B}\ }\textbf {\bibinfo {volume} {89}},\ \bibinfo
  {pages} {220502(R)} (\bibinfo {year} {2014}{\natexlab{a}})}\BibitemShut
  {NoStop}%
\bibitem [{\citenamefont {Mandal}\ \emph
  {et~al.}(2014{\natexlab{b}})\citenamefont {Mandal}, \citenamefont {Cohen},\
  and\ \citenamefont {Haule}}]{PhysRevB.90.060501}%
  \BibitemOpen
  \bibfield  {author} {\bibinfo {author} {\bibfnamefont {S.}~\bibnamefont
  {Mandal}}, \bibinfo {author} {\bibfnamefont {R.~E.}\ \bibnamefont {Cohen}}, \
  and\ \bibinfo {author} {\bibfnamefont {K.}~\bibnamefont {Haule}},\ }\href
  {\doibase 10.1103/PhysRevB.90.060501} {\bibfield  {journal} {\bibinfo
  {journal} {Phys. Rev. B}\ }\textbf {\bibinfo {volume} {90}},\ \bibinfo
  {pages} {060501(R)} (\bibinfo {year} {2014}{\natexlab{b}})}\BibitemShut
  {NoStop}%
\bibitem [{\citenamefont {Mandal}\ \emph {et~al.}(2017)\citenamefont {Mandal},
  \citenamefont {Zhang}, \citenamefont {Ismail-Beigi},\ and\ \citenamefont
  {Haule}}]{PhysRevLett.119.067004}%
  \BibitemOpen
  \bibfield  {author} {\bibinfo {author} {\bibfnamefont {S.}~\bibnamefont
  {Mandal}}, \bibinfo {author} {\bibfnamefont {P.}~\bibnamefont {Zhang}},
  \bibinfo {author} {\bibfnamefont {S.}~\bibnamefont {Ismail-Beigi}}, \ and\
  \bibinfo {author} {\bibfnamefont {K.}~\bibnamefont {Haule}},\ }\href
  {\doibase 10.1103/PhysRevLett.119.067004} {\bibfield  {journal} {\bibinfo
  {journal} {Phys. Rev. Lett.}\ }\textbf {\bibinfo {volume} {119}},\ \bibinfo
  {pages} {067004} (\bibinfo {year} {2017})}\BibitemShut {NoStop}%
\bibitem [{\citenamefont {Kunes}\ \emph {et~al.}(2008)\citenamefont {Kunes},
  \citenamefont {Lukoyanov}, \citenamefont {Anisimov}, \citenamefont
  {Scalettar},\ and\ \citenamefont {Pickett}}]{nmat_kunes}%
  \BibitemOpen
  \bibfield  {author} {\bibinfo {author} {\bibfnamefont {J.}~\bibnamefont
  {Kunes}}, \bibinfo {author} {\bibfnamefont {A.~V.}\ \bibnamefont
  {Lukoyanov}}, \bibinfo {author} {\bibfnamefont {V.~I.}\ \bibnamefont
  {Anisimov}}, \bibinfo {author} {\bibfnamefont {R.~T.}\ \bibnamefont
  {Scalettar}}, \ and\ \bibinfo {author} {\bibfnamefont {W.~E.}\ \bibnamefont
  {Pickett}},\ }\href@noop {} {\bibfield  {journal} {\bibinfo  {journal} {Nat.
  Mater}\ }\textbf {\bibinfo {volume} {7}},\ \bibinfo {pages} {198 } (\bibinfo
  {year} {2008})}\BibitemShut {NoStop}%
\bibitem [{\citenamefont {Haule}\ \emph {et~al.}(2010)\citenamefont {Haule},
  \citenamefont {Yee},\ and\ \citenamefont {Kim}}]{nominal1}%
  \BibitemOpen
  \bibfield  {author} {\bibinfo {author} {\bibfnamefont {K.}~\bibnamefont
  {Haule}}, \bibinfo {author} {\bibfnamefont {C.-H.}\ \bibnamefont {Yee}}, \
  and\ \bibinfo {author} {\bibfnamefont {K.}~\bibnamefont {Kim}},\ }\href
  {\doibase 10.1103/PhysRevB.81.195107} {\bibfield  {journal} {\bibinfo
  {journal} {Phys. Rev. B}\ }\textbf {\bibinfo {volume} {81}},\ \bibinfo
  {pages} {195107} (\bibinfo {year} {2010})}\BibitemShut {NoStop}%
\bibitem [{\citenamefont {Blaha}\ \emph {et~al.}(2001)\citenamefont {Blaha},
  \citenamefont {Schwarz}, \citenamefont {Madsen}, \citenamefont {Kvasnicka},\
  and\ \citenamefont {Luitz}}]{wien2k}%
  \BibitemOpen
  \bibfield  {author} {\bibinfo {author} {\bibfnamefont {P.}~\bibnamefont
  {Blaha}}, \bibinfo {author} {\bibfnamefont {K.}~\bibnamefont {Schwarz}},
  \bibinfo {author} {\bibfnamefont {G.}~\bibnamefont {Madsen}}, \bibinfo
  {author} {\bibfnamefont {D.}~\bibnamefont {Kvasnicka}}, \ and\ \bibinfo
  {author} {\bibfnamefont {J.}~\bibnamefont {Luitz}},\ }\href@noop {} {\emph
  {\bibinfo {title} {An augmented plane wave plus local orbitals program for
  calculating crystal properties}}},\ edited by\ \bibinfo {editor}
  {\bibfnamefont {K.}~\bibnamefont {Schwarz}}\ (\bibinfo  {publisher} {Vienna
  University of Technology, Austria, 2001},\ \bibinfo {year}
  {2001})\BibitemShut {NoStop}%
\bibitem [{\citenamefont {Wu}\ and\ \citenamefont
  {Cohen}(2006)}]{PhysRevB.73.235116}%
  \BibitemOpen
  \bibfield  {author} {\bibinfo {author} {\bibfnamefont {Z.}~\bibnamefont
  {Wu}}\ and\ \bibinfo {author} {\bibfnamefont {R.~E.}\ \bibnamefont {Cohen}},\
  }\href {\doibase 10.1103/PhysRevB.73.235116} {\bibfield  {journal} {\bibinfo
  {journal} {Phys. Rev. B}\ }\textbf {\bibinfo {volume} {73}},\ \bibinfo
  {pages} {235116} (\bibinfo {year} {2006})}\BibitemShut {NoStop}%
\bibitem [{\citenamefont {Kutepov}\ \emph {et~al.}(2010)\citenamefont
  {Kutepov}, \citenamefont {Haule}, \citenamefont {Savrasov},\ and\
  \citenamefont {Kotliar}}]{Kutepov:2010bu}%
  \BibitemOpen
  \bibfield  {author} {\bibinfo {author} {\bibfnamefont {A.}~\bibnamefont
  {Kutepov}}, \bibinfo {author} {\bibfnamefont {K.}~\bibnamefont {Haule}},
  \bibinfo {author} {\bibfnamefont {S.~Y.}\ \bibnamefont {Savrasov}}, \ and\
  \bibinfo {author} {\bibfnamefont {G.}~\bibnamefont {Kotliar}},\ }\href@noop
  {} {\bibfield  {journal} {\bibinfo  {journal} {Phys. Rev. B}\ }\textbf
  {\bibinfo {volume} {82}},\ \bibinfo {pages} {045105} (\bibinfo {year}
  {2010})}\BibitemShut {NoStop}%
\bibitem [{\citenamefont {Czy\ifmmode~\dot{z}\else \.{z}\fi{}yk}\ and\
  \citenamefont {Sawatzky}(1994)}]{FLL}%
  \BibitemOpen
  \bibfield  {author} {\bibinfo {author} {\bibfnamefont {M.~T.}\ \bibnamefont
  {Czy\ifmmode~\dot{z}\else \.{z}\fi{}yk}}\ and\ \bibinfo {author}
  {\bibfnamefont {G.~A.}\ \bibnamefont {Sawatzky}},\ }\href {\doibase
  10.1103/PhysRevB.49.14211} {\bibfield  {journal} {\bibinfo  {journal} {Phys.
  Rev. B}\ }\textbf {\bibinfo {volume} {49}},\ \bibinfo {pages} {14211}
  (\bibinfo {year} {1994})}\BibitemShut {NoStop}%
\bibitem [{\citenamefont {Haule}(2015)}]{exact}%
  \BibitemOpen
  \bibfield  {author} {\bibinfo {author} {\bibfnamefont {K.}~\bibnamefont
  {Haule}},\ }\href {\doibase 10.1103/PhysRevLett.115.196403} {\bibfield
  {journal} {\bibinfo  {journal} {Phys. Rev. Lett.}\ }\textbf {\bibinfo
  {volume} {115}},\ \bibinfo {pages} {196403} (\bibinfo {year}
  {2015})}\BibitemShut {NoStop}%
\bibitem [{url()}]{url1}%
  \BibitemOpen
  \href@noop {} {\ }\bibinfo {note} {Rutgers DFT+DMFT software:
  http:/hauleweb.rutgers.edu/tutorials}\BibitemShut {NoStop}%
\bibitem [{\citenamefont {Gonze}\ \emph {et~al.}(2009)\citenamefont {Gonze},
  \citenamefont {Amadon}, \citenamefont {Anglade}, \citenamefont {Beuken},
  \citenamefont {Bottin}, \citenamefont {Boulanger}, \citenamefont {Bruneval},
  \citenamefont {Caliste}, \citenamefont {Caracas}, \citenamefont {Côté},
  \citenamefont {Deutsch}, \citenamefont {Genovese}, \citenamefont {Ghosez},
  \citenamefont {Giantomassi}, \citenamefont {Goedecker}, \citenamefont
  {Hamann}, \citenamefont {Hermet}, \citenamefont {Jollet}, \citenamefont
  {Jomard}, \citenamefont {Leroux}, \citenamefont {Mancini}, \citenamefont
  {Mazevet}, \citenamefont {Oliveira}, \citenamefont {Onida}, \citenamefont
  {Pouillon}, \citenamefont {Rangel}, \citenamefont {Rignanese}, \citenamefont
  {Sangalli}, \citenamefont {Shaltaf}, \citenamefont {Torrent}, \citenamefont
  {Verstraete}, \citenamefont {Zerah},\ and\ \citenamefont
  {Zwanziger}}]{Gonze20092582}%
  \BibitemOpen
  \bibfield  {author} {\bibinfo {author} {\bibfnamefont {X.}~\bibnamefont
  {Gonze}}, \bibinfo {author} {\bibfnamefont {B.}~\bibnamefont {Amadon}},
  \bibinfo {author} {\bibfnamefont {P.-M.}\ \bibnamefont {Anglade}}, \bibinfo
  {author} {\bibfnamefont {J.-M.}\ \bibnamefont {Beuken}}, \bibinfo {author}
  {\bibfnamefont {F.}~\bibnamefont {Bottin}}, \bibinfo {author} {\bibfnamefont
  {P.}~\bibnamefont {Boulanger}}, \bibinfo {author} {\bibfnamefont
  {F.}~\bibnamefont {Bruneval}}, \bibinfo {author} {\bibfnamefont
  {D.}~\bibnamefont {Caliste}}, \bibinfo {author} {\bibfnamefont
  {R.}~\bibnamefont {Caracas}}, \bibinfo {author} {\bibfnamefont
  {M.}~\bibnamefont {Côté}}, \bibinfo {author} {\bibfnamefont
  {T.}~\bibnamefont {Deutsch}}, \bibinfo {author} {\bibfnamefont
  {L.}~\bibnamefont {Genovese}}, \bibinfo {author} {\bibfnamefont
  {P.}~\bibnamefont {Ghosez}}, \bibinfo {author} {\bibfnamefont
  {M.}~\bibnamefont {Giantomassi}}, \bibinfo {author} {\bibfnamefont
  {S.}~\bibnamefont {Goedecker}}, \bibinfo {author} {\bibfnamefont
  {D.}~\bibnamefont {Hamann}}, \bibinfo {author} {\bibfnamefont
  {P.}~\bibnamefont {Hermet}}, \bibinfo {author} {\bibfnamefont
  {F.}~\bibnamefont {Jollet}}, \bibinfo {author} {\bibfnamefont
  {G.}~\bibnamefont {Jomard}}, \bibinfo {author} {\bibfnamefont
  {S.}~\bibnamefont {Leroux}}, \bibinfo {author} {\bibfnamefont
  {M.}~\bibnamefont {Mancini}}, \bibinfo {author} {\bibfnamefont
  {S.}~\bibnamefont {Mazevet}}, \bibinfo {author} {\bibfnamefont
  {M.}~\bibnamefont {Oliveira}}, \bibinfo {author} {\bibfnamefont
  {G.}~\bibnamefont {Onida}}, \bibinfo {author} {\bibfnamefont
  {Y.}~\bibnamefont {Pouillon}}, \bibinfo {author} {\bibfnamefont
  {T.}~\bibnamefont {Rangel}}, \bibinfo {author} {\bibfnamefont {G.-M.}\
  \bibnamefont {Rignanese}}, \bibinfo {author} {\bibfnamefont {D.}~\bibnamefont
  {Sangalli}}, \bibinfo {author} {\bibfnamefont {R.}~\bibnamefont {Shaltaf}},
  \bibinfo {author} {\bibfnamefont {M.}~\bibnamefont {Torrent}}, \bibinfo
  {author} {\bibfnamefont {M.}~\bibnamefont {Verstraete}}, \bibinfo {author}
  {\bibfnamefont {G.}~\bibnamefont {Zerah}}, \ and\ \bibinfo {author}
  {\bibfnamefont {J.}~\bibnamefont {Zwanziger}},\ }\href {\doibase
  10.1016/j.cpc.2009.07.007} {\bibfield  {journal} {\bibinfo  {journal}
  {Computer Physics Communications}\ }\textbf {\bibinfo {volume} {180}},\
  \bibinfo {pages} {2582 } (\bibinfo {year} {2009})},\ \bibinfo {note} {40
  YEARS OF CPC: A celebratory issue focused on quality software for high
  performance, grid and novel computing architectures}\BibitemShut {NoStop}%
\bibitem [{\citenamefont {Gonze}\ \emph {et~al.}(2005)\citenamefont {Gonze},
  \citenamefont {Rignanese}, \citenamefont {Verstraete}, \citenamefont
  {Betiken}, \citenamefont {Pouillon}, \citenamefont {Caracas}, \citenamefont
  {Jollet}, \citenamefont {Torrent}, \citenamefont {Zerah}, \citenamefont
  {Mikami}, \citenamefont {Ghosez}, \citenamefont {Veithen}, \citenamefont
  {Raty}, \citenamefont {Olevano}, \citenamefont {Bruneval}, \citenamefont
  {Reining}, \citenamefont {Godby}, \citenamefont {Onida}, \citenamefont
  {Hamann},\ and\ \citenamefont {Allan}}]{gonze2005brief}%
  \BibitemOpen
  \bibfield  {author} {\bibinfo {author} {\bibfnamefont {X.}~\bibnamefont
  {Gonze}}, \bibinfo {author} {\bibfnamefont {G.}~\bibnamefont {Rignanese}},
  \bibinfo {author} {\bibfnamefont {M.}~\bibnamefont {Verstraete}}, \bibinfo
  {author} {\bibfnamefont {J.}~\bibnamefont {Betiken}}, \bibinfo {author}
  {\bibfnamefont {Y.}~\bibnamefont {Pouillon}}, \bibinfo {author}
  {\bibfnamefont {R.}~\bibnamefont {Caracas}}, \bibinfo {author} {\bibfnamefont
  {F.}~\bibnamefont {Jollet}}, \bibinfo {author} {\bibfnamefont
  {M.}~\bibnamefont {Torrent}}, \bibinfo {author} {\bibfnamefont
  {G.}~\bibnamefont {Zerah}}, \bibinfo {author} {\bibfnamefont
  {M.}~\bibnamefont {Mikami}}, \bibinfo {author} {\bibfnamefont
  {P.}~\bibnamefont {Ghosez}}, \bibinfo {author} {\bibfnamefont
  {M.}~\bibnamefont {Veithen}}, \bibinfo {author} {\bibfnamefont {J.-Y.}\
  \bibnamefont {Raty}}, \bibinfo {author} {\bibfnamefont {V.}~\bibnamefont
  {Olevano}}, \bibinfo {author} {\bibfnamefont {F.}~\bibnamefont {Bruneval}},
  \bibinfo {author} {\bibfnamefont {L.}~\bibnamefont {Reining}}, \bibinfo
  {author} {\bibfnamefont {R.}~\bibnamefont {Godby}}, \bibinfo {author}
  {\bibfnamefont {G.}~\bibnamefont {Onida}}, \bibinfo {author} {\bibfnamefont
  {D.}~\bibnamefont {Hamann}}, \ and\ \bibinfo {author} {\bibfnamefont
  {D.}~\bibnamefont {Allan}},\ }\href@noop {} {\bibfield  {journal} {\bibinfo
  {journal} {Zeitschrift f{\"u}r Kristallographie.(Special issue on
  Computational Crystallography)}\ }\textbf {\bibinfo {volume} {220}},\
  \bibinfo {pages} {558} (\bibinfo {year} {2005})}\BibitemShut {NoStop}%
\bibitem [{\citenamefont {Haule}\ \emph {et~al.}(2014)\citenamefont {Haule},
  \citenamefont {Birol},\ and\ \citenamefont {Kotliar}}]{nominal2}%
  \BibitemOpen
  \bibfield  {author} {\bibinfo {author} {\bibfnamefont {K.}~\bibnamefont
  {Haule}}, \bibinfo {author} {\bibfnamefont {T.}~\bibnamefont {Birol}}, \ and\
  \bibinfo {author} {\bibfnamefont {G.}~\bibnamefont {Kotliar}},\ }\href
  {\doibase 10.1103/PhysRevB.90.075136} {\bibfield  {journal} {\bibinfo
  {journal} {Phys. Rev. B}\ }\textbf {\bibinfo {volume} {90}},\ \bibinfo
  {pages} {075136} (\bibinfo {year} {2014})}\BibitemShut {NoStop}%
\bibitem [{cou()}]{coulomb2}%
  \BibitemOpen
  \href {http://hauleweb.rutgers.edu/tutorials/CoulombUexplain.html} {\
  }\bibinfo {note}
  {Http://hauleweb.rutgers.edu/tutorials/CoulombUexplain.html}\BibitemShut
  {NoStop}%
\bibitem [{\citenamefont {Georges}\ \emph {et~al.}(2013)\citenamefont
  {Georges}, \citenamefont {de' Medici},\ and\ \citenamefont
  {Mravlje}}]{coulomb}%
  \BibitemOpen
  \bibfield  {author} {\bibinfo {author} {\bibfnamefont {A.}~\bibnamefont
  {Georges}}, \bibinfo {author} {\bibfnamefont {L.}~\bibnamefont {de' Medici}},
  \ and\ \bibinfo {author} {\bibfnamefont {J.}~\bibnamefont {Mravlje}},\ }\href
  {\doibase 10.1146/annurev-conmatphys-020911-125045} {\bibfield  {journal}
  {\bibinfo  {journal} {Annual Review of Condensed Matter Physics}\ }\textbf
  {\bibinfo {volume} {4}},\ \bibinfo {pages} {137} (\bibinfo {year}
  {2013})}\BibitemShut {NoStop}%
\bibitem [{\citenamefont {Okawa}\ \emph {et~al.}(2010)\citenamefont {Okawa},
  \citenamefont {Matsunami}, \citenamefont {Ishizaka}, \citenamefont {Eguchi},
  \citenamefont {Taguchi}, \citenamefont {Chainani}, \citenamefont {Takata},
  \citenamefont {Yabashi}, \citenamefont {Tamasaku}, \citenamefont {Nishino},
  \citenamefont {Ishikawa}, \citenamefont {Kuga}, \citenamefont {Horie},
  \citenamefont {Nakatsuji},\ and\ \citenamefont
  {Shin}}]{PhysRevLett.104.247201}%
  \BibitemOpen
  \bibfield  {author} {\bibinfo {author} {\bibfnamefont {M.}~\bibnamefont
  {Okawa}}, \bibinfo {author} {\bibfnamefont {M.}~\bibnamefont {Matsunami}},
  \bibinfo {author} {\bibfnamefont {K.}~\bibnamefont {Ishizaka}}, \bibinfo
  {author} {\bibfnamefont {R.}~\bibnamefont {Eguchi}}, \bibinfo {author}
  {\bibfnamefont {M.}~\bibnamefont {Taguchi}}, \bibinfo {author} {\bibfnamefont
  {A.}~\bibnamefont {Chainani}}, \bibinfo {author} {\bibfnamefont
  {Y.}~\bibnamefont {Takata}}, \bibinfo {author} {\bibfnamefont
  {M.}~\bibnamefont {Yabashi}}, \bibinfo {author} {\bibfnamefont
  {K.}~\bibnamefont {Tamasaku}}, \bibinfo {author} {\bibfnamefont
  {Y.}~\bibnamefont {Nishino}}, \bibinfo {author} {\bibfnamefont
  {T.}~\bibnamefont {Ishikawa}}, \bibinfo {author} {\bibfnamefont
  {K.}~\bibnamefont {Kuga}}, \bibinfo {author} {\bibfnamefont {N.}~\bibnamefont
  {Horie}}, \bibinfo {author} {\bibfnamefont {S.}~\bibnamefont {Nakatsuji}}, \
  and\ \bibinfo {author} {\bibfnamefont {S.}~\bibnamefont {Shin}},\ }\href
  {\doibase 10.1103/PhysRevLett.104.247201} {\bibfield  {journal} {\bibinfo
  {journal} {Phys. Rev. Lett.}\ }\textbf {\bibinfo {volume} {104}},\ \bibinfo
  {pages} {247201} (\bibinfo {year} {2010})}\BibitemShut {NoStop}%
\bibitem [{\citenamefont {Biswas}\ \emph {et~al.}(2014)\citenamefont {Biswas},
  \citenamefont {Salman}, \citenamefont {Neupert}, \citenamefont {Morenzoni},
  \citenamefont {Pomjakushina}, \citenamefont {von Rohr}, \citenamefont
  {Conder}, \citenamefont {Balakrishnan}, \citenamefont {Hatnean},
  \citenamefont {Lees}, \citenamefont {Paul}, \citenamefont {Schilling},
  \citenamefont {Baines}, \citenamefont {Luetkens}, \citenamefont {Khasanov},\
  and\ \citenamefont {Amato}}]{topo1}%
  \BibitemOpen
  \bibfield  {author} {\bibinfo {author} {\bibfnamefont {P.~K.}\ \bibnamefont
  {Biswas}}, \bibinfo {author} {\bibfnamefont {Z.}~\bibnamefont {Salman}},
  \bibinfo {author} {\bibfnamefont {T.}~\bibnamefont {Neupert}}, \bibinfo
  {author} {\bibfnamefont {E.}~\bibnamefont {Morenzoni}}, \bibinfo {author}
  {\bibfnamefont {E.}~\bibnamefont {Pomjakushina}}, \bibinfo {author}
  {\bibfnamefont {F.}~\bibnamefont {von Rohr}}, \bibinfo {author}
  {\bibfnamefont {K.}~\bibnamefont {Conder}}, \bibinfo {author} {\bibfnamefont
  {G.}~\bibnamefont {Balakrishnan}}, \bibinfo {author} {\bibfnamefont {M.~C.}\
  \bibnamefont {Hatnean}}, \bibinfo {author} {\bibfnamefont {M.~R.}\
  \bibnamefont {Lees}}, \bibinfo {author} {\bibfnamefont {D.~M.}\ \bibnamefont
  {Paul}}, \bibinfo {author} {\bibfnamefont {A.}~\bibnamefont {Schilling}},
  \bibinfo {author} {\bibfnamefont {C.}~\bibnamefont {Baines}}, \bibinfo
  {author} {\bibfnamefont {H.}~\bibnamefont {Luetkens}}, \bibinfo {author}
  {\bibfnamefont {R.}~\bibnamefont {Khasanov}}, \ and\ \bibinfo {author}
  {\bibfnamefont {A.}~\bibnamefont {Amato}},\ }\href {\doibase
  10.1103/PhysRevB.89.161107} {\bibfield  {journal} {\bibinfo  {journal} {Phys.
  Rev. B}\ }\textbf {\bibinfo {volume} {89}},\ \bibinfo {pages} {161107}
  (\bibinfo {year} {2014})}\BibitemShut {NoStop}%
\bibitem [{\citenamefont {Akintola}\ \emph {et~al.}(2017)\citenamefont
  {Akintola}, \citenamefont {Pal}, \citenamefont {Potma}, \citenamefont {Saha},
  \citenamefont {Wang}, \citenamefont {Paglione},\ and\ \citenamefont
  {Sonier}}]{topo2}%
  \BibitemOpen
  \bibfield  {author} {\bibinfo {author} {\bibfnamefont {K.}~\bibnamefont
  {Akintola}}, \bibinfo {author} {\bibfnamefont {A.}~\bibnamefont {Pal}},
  \bibinfo {author} {\bibfnamefont {M.}~\bibnamefont {Potma}}, \bibinfo
  {author} {\bibfnamefont {S.~R.}\ \bibnamefont {Saha}}, \bibinfo {author}
  {\bibfnamefont {X.~F.}\ \bibnamefont {Wang}}, \bibinfo {author}
  {\bibfnamefont {J.}~\bibnamefont {Paglione}}, \ and\ \bibinfo {author}
  {\bibfnamefont {J.~E.}\ \bibnamefont {Sonier}},\ }\href {\doibase
  10.1103/PhysRevB.95.245107} {\bibfield  {journal} {\bibinfo  {journal} {Phys.
  Rev. B}\ }\textbf {\bibinfo {volume} {95}},\ \bibinfo {pages} {245107}
  (\bibinfo {year} {2017})}\BibitemShut {NoStop}%
\bibitem [{\citenamefont {Linden}\ \emph {et~al.}(2000)\citenamefont {Linden},
  \citenamefont {Yamamoto}, \citenamefont {Karppinen}, \citenamefont
  {Yamauchi},\ and\ \citenamefont {Pietari}}]{doi:10.1063/1.126518}%
  \BibitemOpen
  \bibfield  {author} {\bibinfo {author} {\bibfnamefont {J.}~\bibnamefont
  {Linden}}, \bibinfo {author} {\bibfnamefont {T.}~\bibnamefont {Yamamoto}},
  \bibinfo {author} {\bibfnamefont {M.}~\bibnamefont {Karppinen}}, \bibinfo
  {author} {\bibfnamefont {H.}~\bibnamefont {Yamauchi}}, \ and\ \bibinfo
  {author} {\bibfnamefont {T.}~\bibnamefont {Pietari}},\ }\href@noop {}
  {\bibfield  {journal} {\bibinfo  {journal} {Applied Physics Letters}\
  }\textbf {\bibinfo {volume} {76}},\ \bibinfo {pages} {2925} (\bibinfo {year}
  {2000})}\BibitemShut {NoStop}%
\bibitem [{\citenamefont {Zhang}\ \emph {et~al.}(2014)\citenamefont {Zhang},
  \citenamefont {Ma}, \citenamefont {Dai}, \citenamefont {Zhang}, \citenamefont
  {He}, \citenamefont {Normand},\ and\ \citenamefont
  {Yu}}]{PhysRevB.89.174412}%
  \BibitemOpen
  \bibfield  {author} {\bibinfo {author} {\bibfnamefont {J.}~\bibnamefont
  {Zhang}}, \bibinfo {author} {\bibfnamefont {L.}~\bibnamefont {Ma}}, \bibinfo
  {author} {\bibfnamefont {J.}~\bibnamefont {Dai}}, \bibinfo {author}
  {\bibfnamefont {Y.~P.}\ \bibnamefont {Zhang}}, \bibinfo {author}
  {\bibfnamefont {Z.}~\bibnamefont {He}}, \bibinfo {author} {\bibfnamefont
  {B.}~\bibnamefont {Normand}}, \ and\ \bibinfo {author} {\bibfnamefont
  {W.}~\bibnamefont {Yu}},\ }\href {\doibase 10.1103/PhysRevB.89.174412}
  {\bibfield  {journal} {\bibinfo  {journal} {Phys. Rev. B}\ }\textbf {\bibinfo
  {volume} {89}},\ \bibinfo {pages} {174412} (\bibinfo {year}
  {2014})}\BibitemShut {NoStop}%
\end{thebibliography}%
\end{document}